\documentclass{article}

\usepackage{graphicx}
\usepackage{multirow}
\usepackage{amssymb} 
\usepackage{amsthm}
\usepackage{amsmath}
\usepackage{fullpage}
\usepackage[longnamesfirst]{natbib}
\usepackage{enumerate}
\usepackage[table]{xcolor}
\usepackage{amsfonts, bm, amsmath, color, natbib, rotating, amsthm,  lscape, multicol, epstopdf, verbatim, hyperref}  
\usepackage{authblk}
\usepackage{caption}
\usepackage{subcaption}

\usepackage{setspace}
\theoremstyle{plain} \newtheorem{theorem}{Theorem}   

\theoremstyle{definition}    
\newcommand{\utwi}[1]{\mbox{\boldmath $ #1$}}
\theoremstyle{remark}   

\begin{document}

\newif\ifblinded

\title{A Bayesian Multivariate Functional Dynamic Linear Model}

\ifblinded
\author{}
\else

\author{Daniel R. Kowal, David S. Matteson, and David Ruppert\thanks{Kowal is PhD Candidate, Department of Statistical Science, Cornell University, 301 Malott Hall, Ithaca, NY 14853 (E-mail: \href{mailto:drk92@cornell.edu}{drk92@cornell.edu}). Matteson is Assistant Professor, Department of Statistical Science and ILR School, Cornell University, 1196 Comstock Hall, Ithaca, NY 14853 (E-mail: \href{mailto:matteson@cornell.edu}{matteson@cornell.edu}; Webpage: \url{http://www.stat.cornell.edu/\~matteson/}). Ruppert is Andrew Schultz, Jr. Professor of Engineering,  Department of Statistical Science and School of Operations Research and Information Engineering, Cornell University, 1196 Comstock Hall, Ithaca, NY 14853 (E-mail: \href{mailto:dr24@cornell.edu}{dr24@cornell.edu}; Webpage: \url{http://people.orie.cornell.edu/\~davidr/}). The authors thank the editors and two referees for very helpful comments. We also thank Professor Eve De Rosa and Dr. Vladimir Ljubojevic for providing the LFP data and for their helpful discussions. Financial support from NSF grant AST-1312903 (Kowal and Ruppert) and the Cornell University Institute of Biotechnology and the New York State Division of Science, Technology and Innovation (NYSTAR), a Xerox PARC Faculty Research Award, and NSF grant DMS-1455172 (Matteson)   is gratefully acknowledged.}}

\fi



\maketitle

\large

\vspace{-6mm}

\begin{abstract}
We present a Bayesian approach for modeling multivariate, dependent functional data. To account for the three dominant structural features in the data|{\it functional}, {\it time dependent}, and {\it multivariate} components|we extend hierarchical dynamic linear models  for multivariate time series  to the functional data setting. We also develop Bayesian spline theory in a more general constrained optimization framework. The proposed methods identify a time-invariant functional basis for the functional observations, which is smooth and interpretable, and  can be made common across multivariate observations for additional information sharing. The Bayesian framework permits joint estimation of the model parameters, provides exact inference (up to MCMC error) on specific parameters, and allows generalized dependence structures. Sampling from the posterior distribution is accomplished with an efficient Gibbs sampling algorithm. We illustrate the proposed framework with two applications: (1) multi-economy yield curve data from the recent global recession, and (2) local field potential brain signals in rats, for which we develop a multivariate functional time series approach for multivariate time-frequency analysis.  Supplementary materials, including   \texttt{R} code and the multi-economy yield curve data, are available online.
\end{abstract}


\color{black}

\clearpage

\section{Introduction}
We consider a multivariate time series of functional data. Functional data analysis (FDA) methods are widely applicable, including diverse fields such as economics and finance (e.g., \citealp{FDFM}); brain imaging (e.g., \citealp{staicu2012modeling});   chemometric analysis, speech recognition, and electricity consumption \citep{ferraty2006nonparametric}; and growth curves and environmental monitoring \citep{silverman2005functional}. Methodology for independent and identically distributed (iid) functional data has been well-developed, but in the case of {\it dependent} functional data, the iid methods are not appropriate.  Such dependence is common, and can arise via multiple responses,  temporal and spatial effects, repeated measurements, missing covariates, or simply because of some natural grouping in the data (e.g., \citealp{horvath2012inference}). Here, we consider two distinct sources of dependence: time dependence for time-ordered functional observations and  contemporaneous dependence for multivariate functional observations. 

Suppose  we observe multiple functions $Y_t^{(c)}(\tau)$, $c=1,\ldots,C$, at time points $t=1,\ldots,T$. Such observations have three dominant features:
\begin{enumerate}[(a)]
\item For each $c$ and $t$, $Y_t^{(c)}(\tau)$ is a {\it function} of $\tau \in \mathcal{T}$;
\item For each $c$ and $\tau$, $Y_t^{(c)}(\tau)$ is a {\it time series} for $t=1,\ldots,T$; and
\item For each $t$ and $\tau$, $Y_t^{(c)}(\tau)$ is a {\it multivariate} observation with outcomes $c=1,\ldots,C$.
\end{enumerate}
We assume that $\mathcal{T} \subseteq \mathbb{R}^d$ is compact, and focus on the case $d=1$ in which $\tau$ is a scalar. However, our approach may be adapted to the more general setting.

We consider two diverse applications of multivariate functional time series (MFTS).

\noindent {\bf Multi-Economy Yield Curves:}  Let $Y_t^{(c)}(\tau)$ denote {\it multi-economy yield curves}  
observed on weeks  $t=1,\ldots,T$ for economies $c=1,\ldots,C$, which refer to the Federal Reserve, the Bank of England, the European Central Bank, and the Bank of Canada. For a given currency and level of risk of a debt, the yield curve describes the interest rate as a function of the length of the borrowing period, or time to maturity, $\tau$. 
Yield curves are   important  in a variety of economic and financial applications, such as evaluating economic and monetary conditions, pricing fixed-income securities, generating forward curves, computing inflation premiums, and monitoring business cycles \citep{bolder2004empirical}. We are particularly interested in the relationships among yield curves for the aforementioned globally-influential economies, and in how these relationships vary over time. However, existing FDA methods are inadequate to model the dynamic dependences among and between the yield curves for different economies, such as  contemporaneous dependence, volatility clustering, covariates, and change points. Our approach resolves these inadequacies, and provides useful insights into the interactions among multi-economy yield curves (see Section \ref{yields}).

\noindent {\bf  Multivariate Time-Frequency Analysis:} For multivariate time series, the  periodic behavior of the process is often the primary interest. {\it Time-frequency analysis} is used when this periodic behavior varies over time, which requires consideration of both the time and frequency domains (e.g., \citealp{shumway2000time}). Typical  methods segment the multivariate time series into (overlapping) time bins within which the periodic behavior is approximately stationary; within each bin, standard frequency domain or spectral analysis is performed, which uses the multivariate discrete Fourier transform of the time series to identify  dominant frequencies. Interestingly, although the raw signal in this setting is a multivariate time series, time-frequency analysis produces a MFTS: the multivariate discrete Fourier transform is a {\it function} of frequency $\tau$ for {\it time}  bins $t=1,\ldots,T$, where $c=1,\ldots,C$ index the {\it multivariate} components of the spectrum. We analyze  local field potential (LFP) data collected on rats, which measures the neural activity of local brain regions 
over time 
\citep{vladTalk}.
Our interest is in the time-dependent periodic behavior of these  local brain regions under different stimuli, and in particular the synchronization between brain regions. Our novel MFTS approach to time-frequency analysis provides the necessary multivariate structure and inference|which is unavailable in standard time-frequency analysis|to precisely characterize brain behavior under certain stimuli (see Section \ref{rat}).

To model MFTS, we extend the hierarchical dynamic linear model (DLM) framework of \cite{gamerman1993dynamic} and \cite{westDLM}      for multivariate time series to the functional data setting. For smooth, flexible, and optimal function estimates, we extend Bayesian spline theory to a more general constrained optimization framework, which we apply for parameter identifiability. 
Our constraints are explicit in the posterior distribution via appropriate conditioning of the standard Bayesian spline posterior distribution, and the corresponding posterior mean is the solution to an appropriate optimization problem. 
We implement an efficient Gibbs sampler to obtain samples from the joint posterior distribution, which provides exact (up to MCMC error) inference for any parameters of interest. The proposed hierarchical Bayesian {\it Multivariate Functional Dynamic Linear Model} has greater applicability and utility than related methods. It provides flexible modeling of complex dependence structures among the functional observations, such as time dependence, contemporaneous dependence,  stochastic volatility, covariates, and change points, and can incorporate application-specific prior information.

The paper proceeds as follows. In Section \ref{genmod}, we present our model in its most general form. We develop our (factor loading) curve estimation technique in Section \ref{loadings}. In Section \ref{results}, we apply our model to the two applications discussed above and  interpret  the results. 
\ifblinded
\color{blue} \bf
\fi
The corresponding \texttt{R} code and data files  for the yield curve application are available as supplementary materials. We also  provide the details of our Gibbs sampling algorithm, present MCMC diagnostics for our applications,   and include additional figures in the appendix.

\color{black} \rm

\section{A Multivariate Functional Dynamic Linear Model} \label{genmod}
Suppose we observe functions $Y_t^{(c)}\!\colon \mathcal{T}\rightarrow \mathbb{R}$ at times $t=1,\ldots,T$ for outcomes $c=1,\ldots,C$, where $\mathcal{T}\subseteq \mathbb{R}$ is  compact. 
We   refer to the following model as the {\it Multivariate Functional Dynamic Linear Model} (MFDLM):

\begin{equation}\label{fhdlm}
\begin{cases}
\mathbf{Y}_t(\tau)= \mathbf{F}(\tau)  \utwi{\beta}_t  + \utwi{\epsilon}_{t}(\tau),  & \left[\utwi{\epsilon}_{t}(\tau) \big|\mathbf{E}_t \right]  \stackrel{indep}{\sim} N\left(\mathbf{0}, \mathbf{E}_t \right),\\  
\utwi{\beta}_t = \mathbf{X}_t  \utwi{\theta}_t  + \utwi{\nu}_{t}, & \left[\utwi{\nu}_{t} \big|\mathbf{V}_t\right]   \stackrel{indep}{\sim} N(\mathbf{0}, \mathbf{V}_t),\\
\utwi{\theta}_t = \mathbf{G}_t  \utwi{\theta}_{t-1}  + \utwi{\omega}_{t}, & \left[\utwi{\omega}_{t} \big|\mathbf{W}_t\right]    \stackrel{indep}{\sim} N(\mathbf{0}, \mathbf{W}_t), 
\end{cases} 
\end{equation}
where $\mathbf{Y}_t(\tau)= \left[Y_t^{(1)}(\tau)  , Y_t^{(2)}(\tau), \ldots, Y_t^{(C)}(\tau)\right]'$ is the $C$-dimensional vector of multivariate functional observations at time $t$ evaluated at $\tau \in \mathcal{T}$;
\ifblinded
\color{blue} \bf
\fi
$\mathbf{F}(\tau)$ is the $C\times KC$  block matrix  with $1\times K$ diagonal blocks $\left[f_1^{(c)}(\tau), f_2^{(c)}(\tau), \ldots,f_K^{(c)}(\tau)\right]$ for $c=1,\ldots,C$ of {\it factor loading curves} evaluated at $\tau \in \mathcal{T}$, with $K$  the number of factors per outcome, and zeros elsewhere; 
\color{black} \rm
 $ \utwi{\beta}_t = \left[ \beta_{1,t}^{(1)} , \ldots,\beta_{K, t}^{(1)} , \beta_{1,t}^{(2)},\ldots, \beta_{K,t}^{(C)} \right]'$ is the $KC$-dimensional vector of {\it factors} that serve as the time-dependent weights on the factor loading curves;
 $\mathbf{X}_t$ is the known $KC\times p $ matrix of covariates at time $t$, where $p$ is the total number of covariates;
$\utwi{\theta}_t$ is the $p$-dimensional vector of regression coefficients associated with $\mathbf{X}_t$;
$\mathbf{G}_t$ is the $p\times p$ evolution matrix of the regression coefficients $\utwi{\theta}_t$ at time $t$; and
 $\utwi{\epsilon}_t(\tau)$, $\utwi{\nu}_t, $ and $\utwi{\omega}_t$ are mutually independent error vectors with  variance matrices  $\mathbf{E}_t$, $\mathbf{V}_t$, and $\mathbf{W}_t$, respectively. 
 \ifblinded
\color{blue} \bf
\fi
 We assume conditional independence of $[\utwi{\epsilon}_t(\tau)|\mathbf{E}_t]$ over both $t=1,\ldots, T$  and $\tau \in \mathcal{T}$; however, the latter assumption of independence over $\tau$ may be relaxed.  
\color{black} \rm
 We can immediately obtain a useful submodel of \eqref{fhdlm} by excluding covariates,   $\mathbf{X}_t = \mathbf{I}_{CK\times CK}$,  and removing a level of the hierarchy,  $\mathbf{V}_t = \mathbf{0}_{CK\times CK}$, so that setting $\mathbf{G}_t = \mathbf{G}$ models $\utwi{\beta}_t$  ($=\utwi{\theta}_t$,  almost surely) with a vector autoregression (VAR). 
 

To understand \eqref{fhdlm}, first note that the observation level of the model combines the {\it functional} component $\mathbf{F}(\tau)$ with the  {\it multivariate time series} component $\utwi{\beta}_t$. In scalar notation, we can write the observation level as 
\begin{equation} \label{toplevel}
Y_t^{(c)}(\tau) = \sum_{k=1}^K f_k^{(c)}(\tau)\beta_{k,t}^{(c)} + \epsilon_t^{(c)}(\tau)
\end{equation}
in which $\epsilon_t^{(c)}(\tau)$ are the elements of the vector $\utwi{\epsilon}_t(\tau)$. In our construction,  we can always write the observation level of \eqref{fhdlm}  as \eqref{toplevel}; simplifications for the other levels will depend on the choice of submodel. 
 \ifblinded
\color{blue} \bf
\fi
For model identifiability, we require orthonormality of the factor loading curves:
\begin{equation}\label{orthoCon}
 \int_{\tau\in\mathcal{T}}f_k^{(c)}(\tau) f_j^{(c)}(\tau) \ d \tau = \mathbf{1}\!(k=j)  
 \end{equation}
 for $k,j = 1,\ldots,K$ and all outcomes $c=1,\ldots,C$, where $\mathbf{1}\!(\cdot)$ is the indicator function. In addition, to ensure a unique and interpretable ordering of the factors $\beta_{1,t}^{(c)},\ldots,\beta_{K,t}^{(c)}$  for each outcome $c=1,\ldots,C$, we order the factor loading curves $f_1^{(c)}, \ldots, f_K^{(c)}$ by decreasing smoothness. We discuss our implementation of these constraints in    Sections \ref{bayesiansplines} and \ref{concurves}.
\color{black}\rm

There are three primary interpretations of the model, which provide insight into useful extensions and submodels. 

First, we can view \eqref{toplevel} as a basis expansion of the functional observations $Y_t^{(c)}$, with a (multivariate) time series model for the basis coefficients $\beta_{k,t}^{(c)}$ to account for the additional dependence structures, such as common trends (see Section \ref{comHMM}), stochastic volatility (see Section \ref{svm}), and covariates. Since the identifiability constraint in \eqref{orthoCon} expresses orthonormality  with respect to the $L^2$ inner product, we  can interpret   $\left\{f_1^{(c)}, \ldots, f_K^{(c)}\right\}$ as an orthonormal basis for the functional observations $Y_t^{(c)}$. In contrast to common  basis expansion procedures   that assume  the basis functions are known and only the coefficients need to be estimated (e.g., \citealp{bowsher2008dynamics}),  we allow our basis functions $f_k^{(c)}$ to be estimated from the data. As a result, the $f_k^{(c)}$ will be more closely tailored to the data, which reduces the number of functions $K$ needed to adequately fit the data. Conditional on the $f_k^{(c)}$, we can specify the $\utwi{\beta}_t$- and $\utwi{\theta}_t$-levels of \eqref{fhdlm} to appropriately model the remaining dependence among the $Y_t^{(c)}$. 
Using this  interpretation, we also note that \eqref{fhdlm} may be described as a multivariate dynamic (concurrent) functional linear model, and therefore extends a highly useful model in FDA \citep{cardot1999functional}.

Similarly, we can interpret \eqref{fhdlm} as a dynamic factor analysis, which is a common approach in yield curve modeling (e.g., \citealp{FDFM,SDFM}). Under this interpretation, the $\beta_{k,t}^{(c)}$ are dynamic {\it factors} and the $f_k^{(c)}$ are {\it factor loading curves} (FLCs); we will use this terminology for the remainder of the paper. Compared to a standard factor analysis, \eqref{fhdlm} has two major modifications: the factors $\beta_{k,t}^{(c)}$ are dynamic and therefore have an accompanying (multivariate) time series model, and the $f_k^{(c)}$ are functions rather than vectors.

Naturally, \eqref{fhdlm} has strong connections to a hierarchical DLM. Standard hierarchical DLM algorithms  for sampling $\utwi{\beta}_t$ and $\utwi{\theta}_t$ assume that $\{\mathbf{F}, \mathbf{G}_t, \mathbf{X}_t, \mathbf{E}_t, \mathbf{V}_t, \mathbf{W}_t\}$ is known (e.g., \citealp{durbin2002simple,petris2009dynamic}). Within our Gibbs sampler, we may {\it condition} on this set of parameters, and then use existing DLM algorithms to efficiently sample   $\utwi{\beta}_t$ and   $\utwi{\theta}_t$  with minimal implementation effort. Unconditionally, $\mathbf{F}$ is unknown, but we impose the necessary identifiability constraints; see Section \ref{loadings} for more details. $\mathbf{G}_t$ may be known or unknown depending on the application, but in general it supplies the time series structure of the model (along with the time-dependent error variances): in Section \ref{comHMM}, $\mathbf{G}_t = \mathbf{G}$ is unknown  to allow for data-driven  dependence among the multi-economy yield curves,  and in Section \ref{ratmodsec}, $\mathbf{G}_t =   \mathbf{I}_{CK\times CK}$ is chosen to provide parsimonious time-domain smoothing. 
We assume that $\mathbf{X}_t$ is  known, and may consist  of covariates relevant to each outcome  or can be chosen to provide additional shrinkage of $\utwi{\beta}_t$ through $\utwi{\theta}_t$. Although \cite{{gamerman1993dynamic}} suggest that $\dim(\utwi{\theta}_t) < \dim(\utwi{\beta}_t)$ for strict dimension reduction in the hierarchy, we relax this assumption to allow for covariate information.  Finally, we treat the error variance matrices as unknown, but typically there are simplifications available depending on the application and model choice.  We discuss some examples in Section \ref{results}.

We must also specify a choice for $K$.  In the yield curve application, two natural choices are $K=3$ and $K=4$ for comparison with the common parametric yield curve models: the Nelson-Siegel model  \citep{nelson1987parsimonious} and the Svensson model \citep{svensson1994estimating}, both of which can be expressed as submodels of \eqref{fhdlm}; see \cite{diebold2006forecasting} and   \cite{laurini2010bayesian}. 
 \ifblinded
\color{blue} \bf
\fi
More formally, we can treat $K$ as a parameter and estimate it using reversible jump MCMC methods \citep{green1995reversible}, or select $K$ using marginal likelihood. In particular, since we employ a Gibbs sampler, the marginal likelihood estimation procedure of \cite{chib1995marginal} is convenient for many submodels of \eqref{fhdlm}. For more complex models, DIC provides a less computationally intensive approach than either reversible jump MCMC or marginal likelihood, and is very simple to compute. In the appendix, we discuss a fast procedure based on the singular value decomposition from  our initialization algorithm which can be used to estimate a range of reasonable values for $K$. 

\color{black}\rm

\section{Estimating the Factor Loading Curves}\label{loadings}
We would like to model the FLCs $f_k^{(c)}$ in a smooth, flexible, and computationally appealing manner. Clearly, the latter two attributes are important for broader applicability and larger data sets|including larger $T$, larger $C$, and larger $m_t^{(c)}$, where $m_t^{(c)}$ denotes the number of observation points for outcome $c$ at time $t$. The smoothness requirement is fundamental as well: as documented in  \cite{SDFM}, smoothness constraints can  improve forecasting, despite the small biases imposed by such constraints. Smooth curves also tend to be more interpretable, since gradual trends are usually easier to explain than sharp changes or discontinuities.

However, there are some additional complications. First, we must incorporate the identifiability constraints, preferably without severely detracting from the smoothness and goodness-of-fit of the FLCs. We also have  $K$ curves to estimate for each outcome|or perhaps $K$ curves common to all outcomes (see Section \ref{comcurve})|similar to the varying-coefficients model of \cite{hastie1993varying},  conditional on the factors $\beta_{k,t}^{(c)}$. Finally, the observation points for the functions $Y_t^{(c)}$ are likely different for each outcome $c$, and may also vary with time $t$.

\subsection{Splines}\label{splines}
A common approach in nonparametric and semiparametric regression is to express each unknown function $f_k^{(c)}$ as a linear combination of known basis functions, and then estimate the associated coefficients by maximizing a (penalized) likelihood  (e.g., \citealp{wahba1990spline,eubank1999nonparametric,ruppert2003semiparametric}).  We  use B-spline basis functions for their numerical properties and easy implementation, but our methods can accommodate other bases as well. For now, we  ignore dependence on $c$ for notational convenience; this also corresponds to either the univariate case $(C=1)$  or $C >1$ with $\mathbf{E}_t$ diagonal and the FLCs assumed to be {\it a priori} independent for $c=1,\ldots,C$ (see Section \ref{comcurve} for an important alternative).  Following \cite{wand2008semiparametric}, we use cubic splines and the knot sequence $a = \kappa_1 = \ldots = \kappa_4 < \kappa_5 < \ldots < \kappa_{M+4}  < \kappa_{M+5} = \ldots =\kappa_{M+8}= b$, with  $\utwi{\phi}_B = (\phi_{1}, \ldots, \phi_{{M+4}})$  the associated cubic B-spline basis, $M$ the number of interior knots, and $\mathcal{T} = [a,b]$. While we could allow each $f_k$  to have its own B-spline basis and accompanying sequence of knots, there is no obvious reason to do so. 
 \ifblinded
\color{blue} \bf
\fi
In our applications, we use $M=20$ interior knots. 
\color{black} \rm
For knot placement, we prefer a quantile-based approach such as the default method described in \cite{ruppert2003semiparametric}, which is responsive to the location of observation points in the data yet is computationally inexpensive; however, equally-spaced knots may be preferable in some applications.

Explicitly, we write $f_k(\tau)= \utwi{\phi}_B'(\tau) \utwi{d}_k$, where $\utwi{d}_k$ is the $(M+4)$-dimensional vector of unknown coefficients. Therefore, the function estimation problem is reduced to a vector estimation problem. In classical nonparametric regression,  $\utwi{d}_k$ is estimated by maximizing a penalized likelihood, or equivalently solving
\begin{equation}\label{penlike}
\min_{\utwi{d}_k}  -2\log [\mathbf{Y} | \utwi{d}_k]  + \lambda_k\mathcal{P}(\utwi{d}_k)
\end{equation}
where $ [\mathbf{Y} | \utwi{d}_k]$ is a likelihood, $\mathcal{P}$ is a convex penalty function, and $\lambda_k \ge 0$. We express \eqref{penlike} as a log-likelihood multiplied by $-2$ so that for a Gaussian likelihood, \eqref{penlike} is simply a penalized least squares objective.  For greater generality, we leave the likelihood  unspecified, but later consider the likelihood of model \eqref{toplevel}. To penalize roughness, a standard choice for $\mathcal{P}$ is  the $L^2$-norm of the second derivative of $f_k$, which can be written in terms of $\utwi{d}_k$:
\begin{equation}\label{pen}
\mathcal{P}(\utwi{d}_k) = \int_{\tau \in \mathcal{T}} \left[\ddot{f_k}(\tau)\right]^2 d\tau= \utwi{d}_k'\utwi{\Omega}_\phi \utwi{d}_k 
\end{equation}
where $\ddot{f_k}$ denotes the second derivative of $f_k$ and $\utwi{\Omega}_\phi = \int_\mathcal{T} \ddot{\utwi{\phi}_B}{(\tau)}\ddot{\utwi{\phi}_B}'{(\tau)} \ d\tau$, which is easily computable for B-splines. With this choice of penalty,  \eqref{penlike} balances goodness-of-fit with smoothness, where the trade-off is determined by $\lambda_k $.

Since $\mathcal{P}$ is a quadratic in $\utwi{d}_k$, for fixed $\lambda_k$,  \eqref{penlike} is straightforward to solve for many likelihoods, in particular a Gaussian likelihood. Letting $\utwi{\bar{d}}_k$ be this solution, we can estimate $f_k(\tau)$ for any $\tau \in \mathcal{T}$ with $\hat{f}_k(\tau) =\utwi{\phi}_B'(\tau)\utwi{\bar{d}}_k$. For a general knot sequence, the resulting estimator $\hat{f}_k$ is an O'Sullivan spline, or {\it O-spline}, introduced by \cite{o1986statistical} and explored in \cite{wand2008semiparametric}. 
 \ifblinded
\color{blue} \bf
\fi
In the special case of univariate nonparametric regression in which there is a knot at every observation point, $\hat{f}_k$ is a  natural cubic smoothing spline (e.g., \citealp{green1993nonparametric}). 
\color{black}\rm
Alternatively, if we choose a sparser sequence of knots  and set $\lambda_k =0$, $\hat{f}_k$ is a  regression spline (e.g., \citealp{silverman2005functional}). O-splines are numerically stable, possess natural boundary properties, and can be computed efficiently  (cf. \citealp{wand2008semiparametric}).

\subsection{Bayesian Splines}\label{bayesiansplines}
Splines also have a convenient Bayesian interpretation (e.g., \citealp{wahba78,wahba1983bayesian,wahba1990spline,gu1992penalized,van1995splines,berry2002bayesian}). Returning to \eqref{penlike}, we notably have a likelihood term and a penalty term, where the penalty is a function of only   the vector of coefficients $\utwi{d}_k$ and known quantities. Therefore, conditional on $\lambda_k$, the term $\lambda_k \mathcal{P}(\utwi{d}_k)$ provides prior information about $\utwi{d}_k$, for example that   $f_k = \utwi{\phi}_B'\utwi{d}_k$ is smooth. Under this general interpretation, \eqref{penlike} combines the prior information with the likelihood to obtain an estimate of $\utwi{d}_k$.  A natural Bayesian approach is therefore to construct a prior for $\utwi{d}_k$ based on the penalty $\mathcal{P}$, in particular so that the posterior mode of $\utwi{d}_k$ is the solution to \eqref{penlike}. For the most common settings in which the likelihood is Gaussian and the penalty $\mathcal{P}$ is \eqref{pen}, the posterior distribution of $\utwi{d}_k$ will be Gaussian, so the posterior mean will also solve \eqref{penlike}.

 \ifblinded
\color{blue} \bf
\fi
To construct a prior from $\mathcal{P}$, it is computationally and conceptually convenient to reparameterize $\utwi{d}_k$ so that the penalty matrix $\utwi{\Omega}_\phi$ is diagonal. Under a Gaussian prior, this corresponds to prior independence of the components of $\utwi{d}_k$. 
The reparameterization will also affect the basis $\utwi{\phi}_B$, but otherwise will  leave the likelihood in \eqref{penlike} unchanged. Following \cite{wand2008semiparametric}, let $\utwi{\Omega}_\phi = \mathbf{U}_\Omega \mathbf{D}_\Omega \mathbf{U}_\Omega'$ be the singular value  decomposition of $\utwi{\Omega}_\phi$, where $\mathbf{U}_\Omega' \mathbf{U}_\Omega = \mathbf{I}_{(M+4) \times (M+4)}$ and $\mathbf{D}_\Omega$ is a diagonal matrix with $M+2$ positive components. Denote the diagonal matrix of these positive entries by $\mathbf{D}_{\Omega, P}$ and let $\mathbf{U}_{\Omega, P}$ be the corresponding $(M+4) \times (M+2)$ submatrix of $\mathbf{U}_{\Omega}$.  Using the reparameterized basis   $\utwi{\phi}'(\tau) = \left[1, \tau, \utwi{\phi}_B'(\tau)\mathbf{U}_{\Omega, P}\mathbf{D}_{\Omega, P}^{-1/2}\right]$ and penalty $\utwi{d}_k' \utwi{\Omega}_D \utwi{d}_k$ with $\utwi{\Omega}_D = \mbox{diag}\left(0, 0, \lambda_k, \ldots, \lambda_k\right)$,     the new solution $\utwi{\hat{d}}_k$ to \eqref{penlike} satisfies $\hat{f}_k(\tau) = \utwi{\phi}_B(\tau)\utwi{\bar{d}}_k =   \utwi{\phi}'(\tau) \utwi{\hat{d}}_k$; see \cite{wand2008semiparametric} for more details.  It is therefore natural to use the prior $\utwi{d}_k \sim N(\utwi{0},\mathbf{D}_k)$, where $\mathbf{D}_k = \mbox{diag}\left(10^8, 10^8, \lambda_k^{-1}, \ldots, \lambda_k^{-1}\right)$ and $\lambda_k > 0$,  which satisfies $\mathbf{D}_k^{-1} \approx \utwi{\Omega}_D$. Notably, this prior is proper, yet is diffuse over the space of constant and linear functions|which are unpenalized by $\mathcal{P}$. This reparameterization is a common approach for fitting splines using mixed effects model software (e.g., \citealp{ruppert2003semiparametric}).

 \color{black}\rm


Since we assume conditional independence between levels of \eqref{fhdlm}, our conditional likelihood for the FLCs is simply that of model \eqref{toplevel}, but we ignore dependence on $c$ for now: 
\begin{equation}\label{likeli1}
Y_t(\tau) = \sum_{k=1}^K \beta_{k,t}f_k(\tau) + \epsilon_t(\tau) =  \sum_{k=1}^K \beta_{k,t}\utwi{\phi}'(\tau)\utwi{d}_k + \epsilon_t(\tau)
\end{equation}
where $\epsilon_t(\tau) \stackrel{iid}{\sim} N(0, \sigma^2)$ for simplicity; the results are similar for more sophisticated error variance structures. In particular, \eqref{likeli1} describes the distribution of the functional data $Y_t$ given the FLCs $f_k$ (or $\utwi{d}_k$), also conditional on $\beta_{k,t}$ and  $\sigma^2$. 

 Under the likelihood of model \eqref{likeli1} and the reparameterized (approximate) penalty $\utwi{d}_k' \mathbf{D}_k^{-1} \utwi{d}_k$, the solution to \eqref{penlike} conditional on $\utwi{d}_j$, $j\ne k$ is given by $\utwi{\hat{d}}_k = \mathbf{B}_k\mathbf{b}_k$ where $\mathbf{B}_k^{-1} =\mathbf{D}_k^{-1} +  \sigma^{-2}\sum_{t=1}^T \beta_{k,t}^2 \sum_{\tau \in \mathcal{T}_t} \utwi{\phi}(\tau)\utwi{\phi}'(\tau)$, $\mathbf{b}_k = \sigma^{-2} \sum_{t=1}^T \beta_{k,t} \sum_{\tau \in \mathcal{T}_t}  \left[Y_t(\tau)  - \sum_{j\ne k} \beta_{j,t} f_j(\tau)\right]\utwi{\phi}(\tau)$, and $\mathcal{T}_t \subseteq \mathcal{T}$ denotes the discrete set of $|\mathcal{T}_t| = m_t$ observation points for $Y_t$ at time $t$. Note that if $\mathcal{T}_t = \mathcal{T}_1$ for  $t=2,\ldots,T$, then $\mathbf{B}_k$ and $\mathbf{b}_k$ may be rewritten more conveniently in vector notation. Most importantly for our purposes, under the same likelihood induced by \eqref{likeli1} and the prior $\utwi{d}_k \sim N(\utwi{0},\mathbf{D}_k)$, the posterior distribution of $\utwi{d}_k$ is multivariate Gaussian with mean $ \utwi{\hat{d}}_k$ and variance $\mathbf{B}_k$. For convenient computations, \cite{wand2008semiparametric} provide an exact construction of $\utwi{\Omega}_\phi$ and suggest efficient algorithms for $\utwi{\hat{d}}_k$ based on the Cholesky decomposition; we provide more details in the appendix.

 \ifblinded
\color{blue} \bf
\fi

To identify the ordering of the factors and FLCs in \eqref{toplevel}, we constrain the smoothing parameters $\lambda_1 > \lambda_2 > \cdots > \lambda_K > 0 $. While other model constraints are available, this ordering constraint is particularly appealing: it sorts the FLCs $f_k$ by decreasing smoothness, as characterized by the penalty function $\mathcal{P}$,  and leads to a convenient prior distribution on the smoothing parameters $\lambda_k$. In the Bayesian setting, the smoothing parameters   are equivalently the prior precisions of the penalized (nonlinear)  components of $\utwi{d}_k$. Letting $d_{k,j}$ denote the $j$th component of $\utwi{d}_k$, the prior on the FLC basis coefficients is  $d_{k,j} \stackrel{iid}{\sim} N(0, \lambda_k^{-1})$ for $j=3, \ldots, M+4$. This is similar to the hierarchical setting of \cite{gelman2006prior}, in which there are $M+2$ groups for each $\lambda_k, k=1,\ldots,K$. Since $M+2$ is typically large, we follow the \cite{gelman2006prior} recommendation to place  uniform priors on the group standard deviations $\lambda_k^{-1/2}, k=1,\ldots,K$.
Incorporating the ordering constraint, the  conditional priors are $\lambda_{k}^{-1/2} \sim \mbox{Uniform}\left(\ell_{k}, u_{k}\right)$, where $\ell_{1}=0$, $\ell_{k} = \lambda_{k-1}^{-1/2}$ for $k=2,\ldots,K$, $u_{k} =  \lambda_{k+1}^{-1/2}$ for $k=1,\ldots, K-1$, and $u_{K} = 10^4$.  The upper bound on $\lambda_K^{-1/2}$, and therefore all $\lambda_k^{-1/2}$, is chosen to equal   the diffuse prior standard deviation  of   $d_{k,1}$ and $d_{k,2}$. The full conditional distributions of the smoothing parameters $\lambda_k$  are $\mbox{Gamma}\left( \frac{1}{2}(M+1), \frac{1}{2} \sum_{j=3}^{M+4} d_{k, j}^2\right)$ truncated to $(u_k^{-2}, \ell_k^{-2})$ for $k=1,\ldots,K$, where we define $\ell_1^{-2} = \infty$.   
Notably, we avoid  the  diffuse Gamma prior on $\lambda_k$, which can be undesirably informative 
and   is strongly discouraged by \cite{gelman2006prior}. More generally, our approach   provides a natural and data-driven method for estimating the smoothing parameters, yet does not inhibit inference. Details on the   sampling of $\lambda_k$  are provided in the appendix.

\color{black}\rm

\subsection{Constrained Bayesian Splines}\label{concurves}
We  extend the Bayesian spline approach to accommodate the necessary identifiability constraints for the MFDLM.  
 \ifblinded
\color{blue} \bf
\fi
For each $k=1,\ldots,K$,  we impose the orthonormality constraints $\int_\mathcal{T} f_k(\tau)f_j(\tau) = \mathbf{1}(k = j)$  for $j=1,\ldots,K$. 
\color{black}\rm
The unit-norm constraint preserves identifiability with respect to scaling, i.e., relative to the factors $\beta_{k,t}$ (up to  changes in sign).  The   orthogonality constraints distinguish between pairs of FLCs, and in our approach identify the FLCs with distinct posterior distributions.


While other identifiability constraints are available for the $f_k$, orthonormality is appealing for a number of reasons. As discussed in Section \ref{genmod}, the orthonormality constraints suggest  that we can interpret  $\left\{f_1,\ldots,f_K\right\}$ as an orthonormal basis for the functional observations $Y_t$. As such, the orthogonality constraints help  eliminate any information overlap between FLCs, which keeps the total number of necessary FLCs to a minimum. Furthermore, the unit norm constraint allows for easier comparisons among the $f_k$. Of course, the $f_k$ will be weighted by the factors $\beta_{k,t}$, so they can still have varying effects on the conditional mean of ${Y}_t$ in \eqref{toplevel}. Finally,  we can write the constraints conveniently in terms of the vectors $\utwi{d}_k$ and $\utwi{d}_j$:
\begin{equation} \label{ident}
\int_{\tau \in \mathcal{T}} f_k(\tau)f_j(\tau) \ d\tau = \int_{\tau \in \mathcal{T}} \utwi{\phi}'(\tau)\utwi{d}_k \utwi{\phi}'(\tau)\utwi{d}_j \ d\tau = \utwi{d}_k'\mathbf{J}_{\phi} \utwi{d}_j = \mathbf{1}(k = j)
\end{equation}
for $j=1,\ldots, K$,   where $\mathbf{J}_{\phi} = \int_{\tau\in\mathcal{T}} \utwi{\phi}(\tau)\utwi{\phi}'(\tau)\ d\tau$ is easily  computed for B-splines, and only needs to be computed once, prior to any MCMC sampling.

The addition of an orthogonality constraint to a (penalized) least squares problem has an intuitive regression-based interpretation, which we present in the following theorem:
\begin{theorem}\label{thm}
Consider the penalized least squares objective $\sigma^{-2}\sum_{i=1}^n (y_i - \mathbf{X}_i'\utwi{d})^2 +\lambda \utwi{d}'\utwi{\Omega}\utwi{d}$, where $y_i \in \mathbb{R}$, $\utwi{d}$ is an unknown $(M+4)$-dimensional vector, $\mathbf{X}_i$ is a known  $(M+4)$-dimensional vector, $\utwi{\Omega}$ is a known $(M+4)\times (M+4)$ positive-definite matrix, and $\sigma^2, \lambda >0$ are known scalars. The solution is $\utwi{\hat{d}}= \mathbf{Bb}$, where $\mathbf{B}^{-1} = \lambda\utwi{\Omega} + \sigma^{-2}\sum_{i=1}^n\mathbf{X}_i\mathbf{X}_i'  $ and $\mathbf{b} = \sigma^{-2} \sum_{i=1}^n \mathbf{X}_i y_i$. Now consider the same objective, but subject to the $J$ linear constraints $\utwi{d}'\mathbf{L} = \mathbf{0}$ for $\mathbf{L}$ a known $(M+4)\times J$ matrix of rank $J$. The  solution is  $\utwi{\tilde{d}} = \mathbf{B}\mathbf{\tilde{b}}$, where $\mathbf{\tilde{b}}$ is the vector of residuals from the generalized least squares regression $\mathbf{b} = \mathbf{L} \utwi{\Lambda} + \utwi{\delta}$ with $\mathbb{E}(\utwi{\delta}) = 0$ and $\mbox{\rm Var}(\utwi{\delta}) = \mathbf{B}$.
\begin{proof}
The optimality of $\utwi{\hat{d}}$ is a well-known result. For the constrained case, the Lagrangian is  $\mathcal{L}(\utwi{d}, \utwi{\Lambda}) =  \sigma^{-2}\sum_{i=1}^n (y_i - \mathbf{X}_i'\utwi{d})^2 + \lambda \utwi{d}'\utwi{\Omega}\utwi{d}+ \utwi{d}'\mathbf{L}\utwi{\Lambda}$, where $\utwi{\Lambda}$ is the $J$-dimensional vector of Lagrange multipliers associated with  the $J$  linear constraints. It is straightforward to minimize $\mathcal{L}(\utwi{d}, \utwi{\Lambda})$ with respect to $\utwi{d}$ and obtain the solution $\utwi{\tilde{d}} = \mathbf{B}\mathbf{\tilde{b}} = \mathbf{B}(\mathbf{b} - \mathbf{L}\utwi{\Lambda})$. Similarly, solving $\nabla \mathcal{L}(\utwi{\tilde{d}}, \utwi{\Lambda}) = \mathbf{0}$ for $\utwi{\Lambda}$ implies that $\utwi{\Lambda} = (\mathbf{L}'\mathbf{B}\mathbf{L})^{-1}\mathbf{L}'\mathbf{B}\mathbf{b}$, which is the solution to the generalized least squares regression of $\mathbf{b}$ on $\mathbf{L}$ with error variance $\mathbf{B}$.  
\end{proof}
\end{theorem}
The result is interpretable: to incorporate linear constraints into a penalized least squares regression, we find $\mathbf{\tilde{b}}$ nearest to $\mathbf{b}$ under the inner product induced by $\mathbf{B}$ among vectors in the space orthogonal to $\mbox{Col}(\mathbf{L})$. In our setting, extending \eqref{penlike} under a Gaussian likelihood to accommodate the (linear)  orthogonality constraints $\utwi{d}_k ' \mathbf{J}_\phi \utwi{d}_j = 0$ for $j \ne k$ may be described via a regression of the unconstrained solution on the  constraints. However, the unit norm constraint  is nonlinear. This constraint affects the scaling but not the shape of $f_k$. Therefore, a reasonable approach is to construct a posterior distribution for $\utwi{d}_k$ that respects the (linear)  orthogonality constraints only, and then normalize the samples from this  posterior  to preserve identifiability. We provide more details in the appendix.

 \ifblinded
\color{blue} \bf
\fi

To extend the unconstrained Bayesian splines of Section \ref{bayesiansplines} to incorporate the orthogonality constraints, we 
   write the constraints $\utwi{d}_k' \mathbf{J}_\phi \utwi{d}_j = 0$ for $j \ne k$ as the linear constraints in Theorem  \ref{thm} with $\mathbf{L}_{[-k]} =\left( \mathbf{J}_{{\phi}} \utwi{d}_1, \ldots,  \mathbf{J}_{{\phi}}\utwi{d}_{k-1}, \mathbf{J}_{{\phi}}\utwi{d}_{k+1}, \ldots, \mathbf{J}_{{\phi}}\utwi{d}_{K}\right)$ and $J = K-1$.   
   Using the full conditional posterior distribution  $\utwi{d}_k \sim N(\mathbf{B}_k\mathbf{b}_k, \mathbf{B}_k)$ from Section \ref{bayesiansplines}, we can additionally {\it condition}   on the linear constraints $\utwi{d}_k' \mathbf{L}_{[-k]} = \utwi{0}$, and obtain the constrained full conditional distribution $\utwi{d}_k \sim N(\mathbf{\tilde{B}}_k\mathbf{b}_k, \mathbf{\tilde{B}}_k)$, where $\mathbf{\tilde{B}}_k  = \mathbf{B}_k - \mathbf{B}_k \mathbf{L}_{[-k]} (\mathbf{L}_{[-k]}'\mathbf{B}_k\mathbf{L}_{[-k]})^{-1}\mathbf{L}_{[-k]}'\mathbf{B}_k$.   Conditioning on the orthogonality constraints is particularly interpretable in the Bayesian setting, and is convenient for posterior sampling; see the appendix for more details.  By comparison, Theorem \ref{thm} implies that the solution to \eqref{penlike} under the likelihood of model \eqref{likeli1}, the  penalty $\utwi{d}_k' \mathbf{D}_k^{-1} \utwi{d}_k$, and subject to the linear constraints $\utwi{d}_k' \mathbf{L}_{[-k]} = \mathbf{0}$ is given by $\utwi{\tilde{d}}_k = \mathbf{B}_k \mathbf{\tilde{b}}_k$, where $\mathbf{\tilde{b}}_k = \mathbf{b}_k - \mathbf{L}_{[-k]} \utwi{\Lambda}_{[-k]}$ and $\utwi{\Lambda}_{[-k]} = (\mathbf{L}_{[-k]}'\mathbf{B}_k\mathbf{L}_{[-k]})^{-1}\mathbf{L}_{[-k]}'\mathbf{B}_k\mathbf{b}_k$.  Notably,  $\mathbf{\tilde{B}}_k\mathbf{b}_k = \mathbf{B}_k\mathbf{\tilde{b}}_k = \utwi{\tilde{d}}_k$, which is a useful result: by simply conditioning on the linear orthogonality constraints in the full conditional Gaussian distribution for $\utwi{d}_k$, the posterior mean of the resulting Gaussian distribution solves the constrained regression problem of Theorem \ref{thm}. In this sense, the identifiability constraints on $f_k$ are enforced optimally. 
   
   \color{black}\rm

\subsection{Common Factor Loading Curves for Multivariate Modeling}\label{comcurve}
Reintroducing dependence on $c$ for the FLCs  $f_k^{(c)}$, suppose that $C>1$, so that our functional time series $Y_t^{(c)}$ is truly multivariate. If we wish to estimate {\it a priori} independent FLCs for each outcome $c$ (with $\mathbf{E}_t$ diagonal), then we can sample from the relevant posterior distributions independently for $c=1,\ldots,C$ using the methods of Section \ref{concurves}. The more interesting case is the {\it common factor loading curves model} given by $f_k^{(c)} = f_k$, so that all outcomes share a common set of FLCs. In the basis interpretation of the MFDLM, this corresponds to the assumption that the functional observations for all outcomes $Y_t^{(c)}$, $c=1,\ldots,C$, $t=1,\ldots, T$ share a common basis. We find this approach to be useful and intuitive, since it pools information across outcomes and suggests a more parsimonious model. Equally important, the common FLCs approach allows for direct comparison between factors $\beta_{k,t}^{(c)}$ and $\beta_{k,t}^{(c')}$ for outcomes $c$ and $c'$, since these factors serve as weights on the {\it same} FLC (or basis function)  $f_k$. We use this model in both applications in Section \ref{results}.

The common FLCs model implies $f_k^{(c)}(\tau)= \utwi{\phi}_{(c)}'(\tau)\utwi{d}_{k}^{(c)} =f_k(\tau) $. However, since the FLCs for each outcome are identical, it is reasonable to assume that they have the same vector of basis functions $\utwi{\phi}$, so $f_k^{(c)} = f_k$ is equivalent to $\utwi{d}_{k}^{(c)} = \utwi{d}_k$. Moreover, by writing $f_k^{(c)}(\tau) = \utwi{\phi}'(\tau)\utwi{d}_{k}$, we can use all of the observation points across all outcomes $c=1,\ldots,C$ and times $t=1,\ldots,T$, yet the parameter of interest, $\utwi{d}_k$, will only be $(M+4)$-dimensional.

Modifying our previous approach, we use the likelihood of model \eqref{toplevel} with the simple error distribution $\epsilon_t^{(c)}(\tau) \stackrel{iid}{\sim} N (0, \sigma_{(c)}^2)$. The implied full conditional posterior distribution for $\utwi{d}_k$ is again $N(\mathbf{\tilde{B}}_k\mathbf{b}_k, \mathbf{\tilde{B}}_k)$, but now with $\mathbf{B}_k^{-1} =\mathbf{D}_k^{-1} +  \sum_{c=1}^C \sigma_{(c)}^{-2}\sum_{t \in T^{(c)}} (\beta_{k,t}^{(c)})^2\sum_{\tau \in \mathcal{T}_{t}^{(c)}}  \utwi{\phi}(\tau)\utwi{\phi}'(\tau)$ and $\mathbf{b}_k = \sum_{c=1}^C \sigma_{(c)}^{-2}\sum_{t \in T^{(c)}} \beta_{k,t}^{(c)} \sum_{\tau \in \mathcal{T}_{t}^{(c)}} \left[Y_{t}^{(c)}(\tau)  - \sum_{j\ne k} \beta_{j,t}^{(c)} f_j(\tau)\right]\utwi{\phi}(\tau)$. For full generality, we allow the (discrete) set of times $T^{(c)}$ to vary for each outcome $c$ and the (discrete) set of observation points $\mathcal{T}_t^{(c)}$  to vary with both time $t$ and outcome $c$, with $|\mathcal{T}_t^{(c)}| = m_t^{(c)}$. Note that we reuse the same notation from Section \ref{concurves} to emphasize the similarity of the multivariate results to the univariate (or {\it a priori} independent FLC) results.
The common notation also allows for a more concise description of the sampling algorithm, which we present in the appendix.

\section{Data Analysis and Results}\label{results}

\subsection{Multi-Economy Yield Curves}\label{yields}
\begin{figure}[h]
  \centering
\includegraphics[scale= .3]{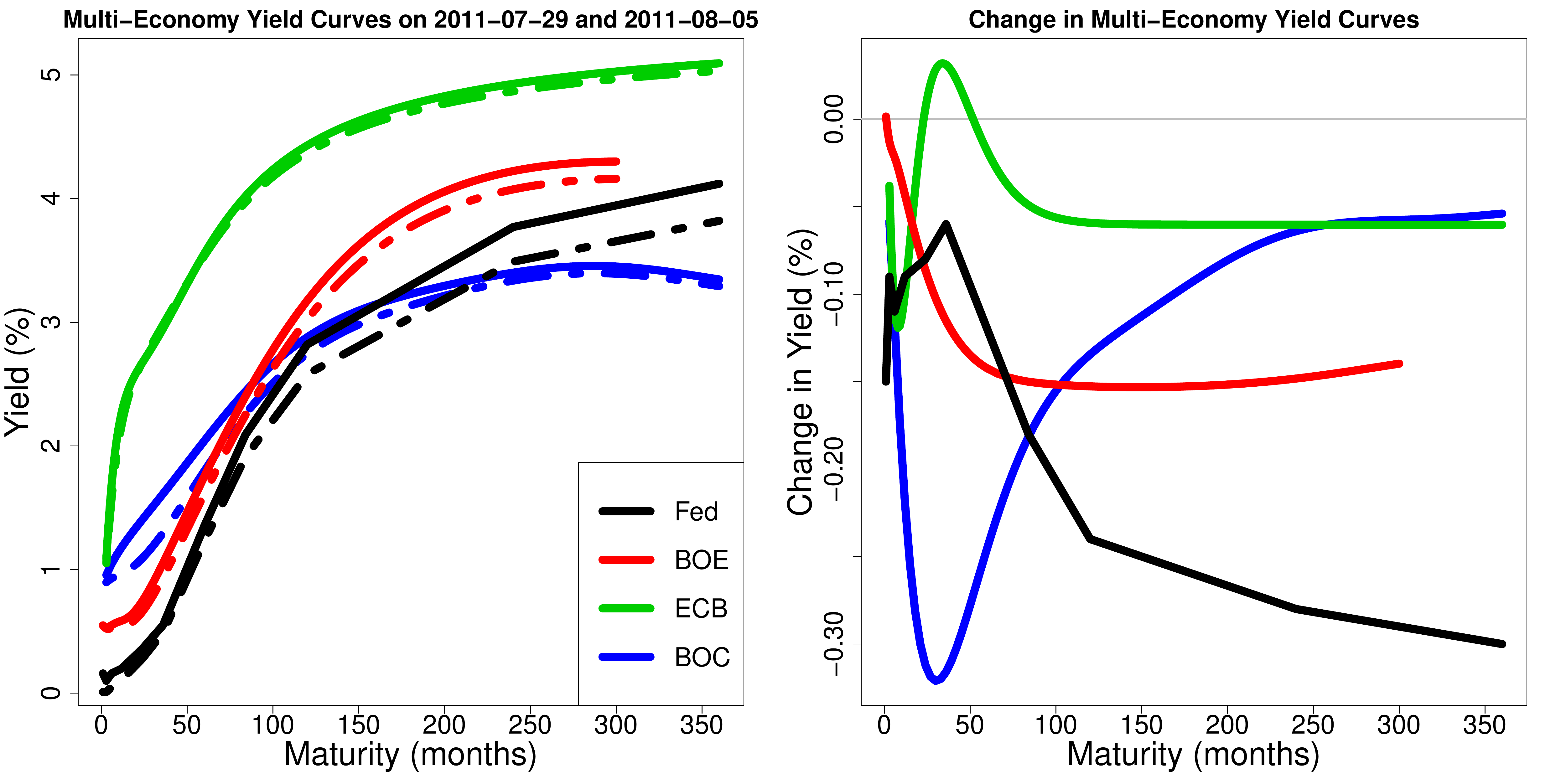} 
  \caption{Multi-economy yield curves from July 29, 2011 (solid) and August 5, 2011 (dashed),  together with the corresponding one-week change curves.}
          \label{fig:yields}
\end{figure}

We jointly analyze {\it weekly} yield curves provided by the Federal Reserve (Fed), the Bank of England (BOE), the European Central Bank (ECB), and the Bank of Canada (BOC; \citealt{bolder2004empirical})  from late 2004 to early 2014 ($T = 490$ and $C=4$). These data are publicly available and published  on the respective central bank websites|and as such, we treat them as reliable estimates of the yield curves. For each outcome, the yield curves are estimated differently: the Fed uses quasi-cubic splines, the BOE uses cubic splines with variable smoothing parameters  \citep{waggoner1997spline}, the ECB uses Svensson curves, and the BOC uses exponential  splines \citep{li2001merrill}. 
Therefore, the functional observations have already been smoothed, although by different procedures. The  available set of maturities $\mathcal{T}_t^{(c)}$  is not the same across economies $c$, and occasionally varies with time $t$. The most frequent values of $m_t^{(c)}$, $t=1,\ldots,T$, are 11 (Fed), 100 (BOE), 354 (ECB), and 120 (BOC), with maturities $\tau$ ranging from 1-3 months up to 300-360 months.  
 \ifblinded
\color{blue} \bf
\fi
To facilitate a simpler analysis, we let $Y_t^{(c)}(\tau)$ be the week-to-week {\it change} in the $c$th central bank yield curve on week $t$ for maturity $\tau$. Differencing the yield curves conveniently addresses the nonstationarity in the weekly data,  and, because the yield curves are pre-smoothed, does not introduce any  notable difficulties with time-varying observation points. We show an example of the multi-economy yield curves  observed at adjacent times on July 29, 2011 and August 5, 2011, as well as the corresponding one-week change in Figure \ref{fig:yields}.

\color{black}\rm

The literature on yield curve modeling is extensive.  Yield curve models commonly adopt the Nelson-Siegel parameterization \citep{nelson1987parsimonious}, often within a state space framework (e.g., \citealp{diebold2006forecasting,diebold2006macroeconomy,diebold2008global,koopman2010analyzing}). Many Bayesian models  also use the Nelson-Siegel or Svensson parameterizations (e.g., \citealp{laurini2010bayesian,cruz2011estimating}). However, the Nelson-Siegel parameterization does not extend to other applications, and often requires solving computationally intensive nonlinear optimization problems. More similar to our approach are the Functional Dynamic Factor Model (FDFM) of \cite{FDFM} and the Smooth Dynamic Factor Model (SDFM)  of \cite{SDFM}, both of which feature nonparametric functional components within a state space framework. The FDFM cleverly uses an EM algorithm to jointly estimate the functional and time series components of the model. However, the EM algorithm makes more sophisticated (multivariate) time series models more challenging to implement, and introduces some difficulties with generalized cross-validation (GCV) for estimation of the nonparametric smoothing parameters. The SDFM avoids GCV and instead  relies on hypothesis tests to select the number and location of knots|and therefore determine the smoothness of the curves. However, this suggests that the smoothness of the curves depends on the significance levels used for the hypothesis tests, of which there can be a substantial number as $m_t^{(c)}$, $C$, or $T$ grow large. By comparison, our smoothing parameters naturally depend on the data through the posterior distribution, which notably does   {\it not} create any difficulties for inference.

The multi-economy yield curves application is a natural setting for the common FLCs model of Section \ref{comcurve}.  First, since $f_k^{(c)} = f_k$ for $c=1,\ldots,C$, the functional component of the MFDLM is the same for all economies, which helps reconcile the aforementioned different central bank yield curve estimation techniques. More specifically, the conditional expectations $\mu_t^{(c)}(\tau) \equiv \sum_{k=1}^K \beta_{k,t}^{(c)} f_k(\tau)$ are linear combinations of the {\it same}  $\left\{f_1,\ldots,f_K\right\}$, and therefore are more directly comparable  for $c=1,\ldots,C$. Second, the common FLCs model is very useful when the set of observed maturities $\mathcal{T}_t^{(c)}$ varies with either outcome $c$ or time $t$. Since the $f_k$ are estimated using {\it all} of the observed maturities $\cup_{t,c} \mathcal{T}_t^{(c)}$, we notably do  not need a missing data model for unobserved maturities at time $t$ for economy $c$. 
In addition, for any $\tau \in \mbox{int range}\left(\cup_{t,c} \mathcal{T}_t^{(c)}\right)$, we may estimate $f_k(\tau)$ and $\mu_t^{(c)}(\tau)$ without any spline-related boundary problems|even when $\tau \not\in \mbox{range}\left(\mathcal{T}_t^{(c)}\right)$. 
By comparison, non-common FLCs|or more generally, any linear combination of outcome-specific natural cubic splines|would impose a linear fit for $\tau \not\in \mbox{range}\left(\mathcal{T}_t^{(c)}\right)$, which may not be reasonable for some applications.

\subsubsection{The Common Trend Model}\label{comHMM}
To investigate the similarities and relationships among  the $C=4$ economy yield curves, we implement the following parsimonious model for multivariate dependence among the factors: 
\begin{equation}\label{commonHMM}
\begin{cases}
\beta_{k,t}^{(1)} = \omega_{k,t}^{(1)}\\
\beta_{k,t}^{(c)} = \gamma_k^{(c)}\beta_{k,t}^{(1)} + \omega_{k,t}^{(c)} & c=2,\ldots,C
\end{cases}
\end{equation}
 \ifblinded
\color{blue} \bf
\fi
where  $\gamma_k^{(c)} \in \mathbb{R}$ is the economy-specific slope term for each factor with the diffuse conjugate prior $\gamma_k^{(c)} \stackrel{iid}{\sim} N(0, 10^8)$. For the errors $\omega_{k,t}^{(c)}$, we use independent AR($r$) models with time-dependent variances, which we discuss in more detail in Section \ref{svm}.  We also implement an interesting extension of \eqref{commonHMM} based on the  autoregressive regime switching models of  \cite{albert1993bayes} and \cite{mcculloch1993bayesian} using the model 
$\beta_{k,t}^{(c)} =  s_{k,t}^{(c)} (\gamma_k^{(c)}\beta_{k,t}^{(1)}) + \omega_{k,t}^{(c)}$, where $\left\{s_{k,t}^{(c)}: t=1,\ldots,T\right\}$ is a discrete Markov chain with states $\{0,1\}$. While this more complex model is not supported by DIC, it is a useful example of the flexibility of the MFDLM; we provide the details in the appendix. 

\color{black}\rm

Letting  $c=1$ correspond  to the Fed yield curve, we can use \eqref{commonHMM} to investigate how the   factors $ \beta_{k,t}^{(c)}$ for each economy $c>1$ are {\it directly} related to those of the Fed, $ \beta_{k,t}^{(1)}$. Since the U.S. economy is commonly regarded as a dominant presence in the global economy (e.g., \citealp{dees2011role}), the Fed yield curve is a natural and interesting reference point. Model \eqref{commonHMM} relates each economy $c>1$ to the Fed using a  regression framework, in which we regress $\beta_{k,t}^{(c)}$ on $\beta_{k,t}^{(1)}$  with AR($r$) errors; since the yield curves were differenced, there is no need (or evidence) for an intercept. 
 \ifblinded
\color{blue} \bf
\fi
The slope parameters $\gamma_k^{(c)}$ measure the strength of this relationship for each factor $k$ and economy $c$. In addition, we can investigate the residuals $\omega_{k,t}^{(c)}$ to determine times $t$ for which $\beta_{k,t}^{(c)}$ deviated substantially from the linear dependence on $\beta_{k,t}^{(1)}$ assumed in model \eqref{commonHMM}. Such periods of uncorrelatedness can offer insight into the interactions between the U.S. and other economies. 
\color{black}\rm

\subsubsection{Stochastic Volatility Models}\label{svm}
For the errors $\omega_{k,t}^{(c)}$ in \eqref{commonHMM}, we use independent AR($r$) models with time-dependent variances, i.e., $\omega_{k,t}^{(c)} = \sum_{i=1}^r \psi_{k, i}^{(c)}\omega_{k,t-i}^{(c)} + \sigma_{k,(c),t} z_{k,t}^{(c)}$ with $z_{k,t}^{(c)} \stackrel{iid}{\sim} N(0,1)$,  $c=1,\ldots,C$.  The AR($r$) specification accounts for the time dependence of the yield curves, while the $\sigma_{k,(c),t}^2$ model the observed volatility clustering.  This latter component is important: in applications of financial time series, it is very common|and often necessary for proper inference|to include a model for the   volatility  (e.g., \citealp{taylor1994modeling,harvey1994multivariate}). It is reasonable to suppose that applications of  financial {\it functional}  time series may also require volatility modeling; the weekly yield curve data provide one such example. Notably,  our hierarchical Bayesian approach  seamlessly incorporates volatility modeling, since, conditional on the volatilities,  DLM algorithms require no additional adjustments for posterior sampling.

Within the Bayesian framework of the MFDLM, it is most natural to use a stochastic volatility model (e.g., \citealp{kim1998stochastic,chib2002markov}). Stochastic volatility models are parsimonious, which is important in hierarchical modeling, yet are highly competitive with more heavily parameterized GARCH models \citep{dan1998}. 
 \ifblinded
\color{blue} \bf
\fi
We model the log-volatility, $\log(\sigma_{(c),k,t}^2)$, as a stationary AR(1) process (for fixed $c$ and $k$), using  the priors and the efficient MCMC sampler of \cite{kastner2014ancillarity}. We provide a plot of the volatilities $\sigma_{k,(c),t}^2$ and additional model details   in the appendix. 

\color{black}\rm



\subsubsection{Results}\label{yieldResults}

 \ifblinded
\color{blue} \bf
\fi
We fit model \eqref{commonHMM} to the multi-economy yield curve data, using the
the \cite{kastner2014ancillarity} model for the volatilities and setting $r=1$, which adequately models the time dependence of the factors, with the diffuse stationarity prior $\psi_{k,1}^{(c)} \stackrel{iid}{\sim}N(0,10^8)$ truncated  to $(-1,1)$. We use the  common FLCs model of Section \ref{comcurve},  and let $\mathbf{E}_t = \mbox{diag}\left(\sigma_{(1)}^2, \ldots, \sigma_{(C)}^2\right)$ with $\sigma_{(c)}^{-2} \stackrel{iid}{\sim} \mbox{Gamma}\left(0.001, 0.001\right)$. We prefer the choice $K =4$, which   corresponds to the number of curves in the Svensson model.  However, since the observations $Y_t^{(c)}$  and the conditional expectations $\mu_t^{(c)}(\tau)$ are both smooth by construction, the errors $\epsilon_t^{(c)}$ are also smooth|and therefore correlated with respect to $\tau$. To mitigate the effects of the error correlation, we increase the number of factors to $K=6$, so that the fitted model \eqref{toplevel}  explains more than 99.5\% of the variability in $Y_t^{(c)}(\tau)$. Since we are primarily interested in the first four factors, we fix $\gamma_k^{(c)} = 0$ for $k > 4$ in model \eqref{commonHMM}, so the two additional factors for each outcome are modeled as independent AR(1) processes with stochastic volatility. We ran the MCMC sampler for $7,000$ iterations and discarded the first $2,000$ iterations as a burn-in. The MCMC sampler is efficient, especially for the factors $\beta_{k,t}^{(c)}$ and the common FLCs $f_k$; we provide the MCMC diagnostics in the appendix.

\color{black}\rm

In Figure \ref{fig:cFLC}, we plot the posterior means of the common FLCs $f_k$ for $k=1,\ldots, 4$. We can interpret these $f_k$ as estimates of the time-invariant underlying functional structure of the yield curves shared by the Fed, the BOE, the ECB, and the BOC. The FLCs are very smooth, and the dominant  hump-like features  occur at different maturities|following from the orthonormality constraints|which allows the model to fit a variety of yield curve shapes. Interestingly, the estimated $f_1,f_2,$ and $f_3$ are similar to the level, slope, and curvature functions of the Nelson-Siegel parameterization described by \cite{diebold2006forecasting}.
Since the factors $\beta_{k,t}^{(c)}$ serve as weights on the FLCs $f_k$ in \eqref{toplevel}, we may interpret the factors $\beta_{k,t}^{(c)}$|and therefore the slopes $\gamma_{k}^{(c)}$|based on these features of the yield curve explained by the corresponding $f_k$.

\begin{figure}[h]
  \centering
\includegraphics[scale= .4]{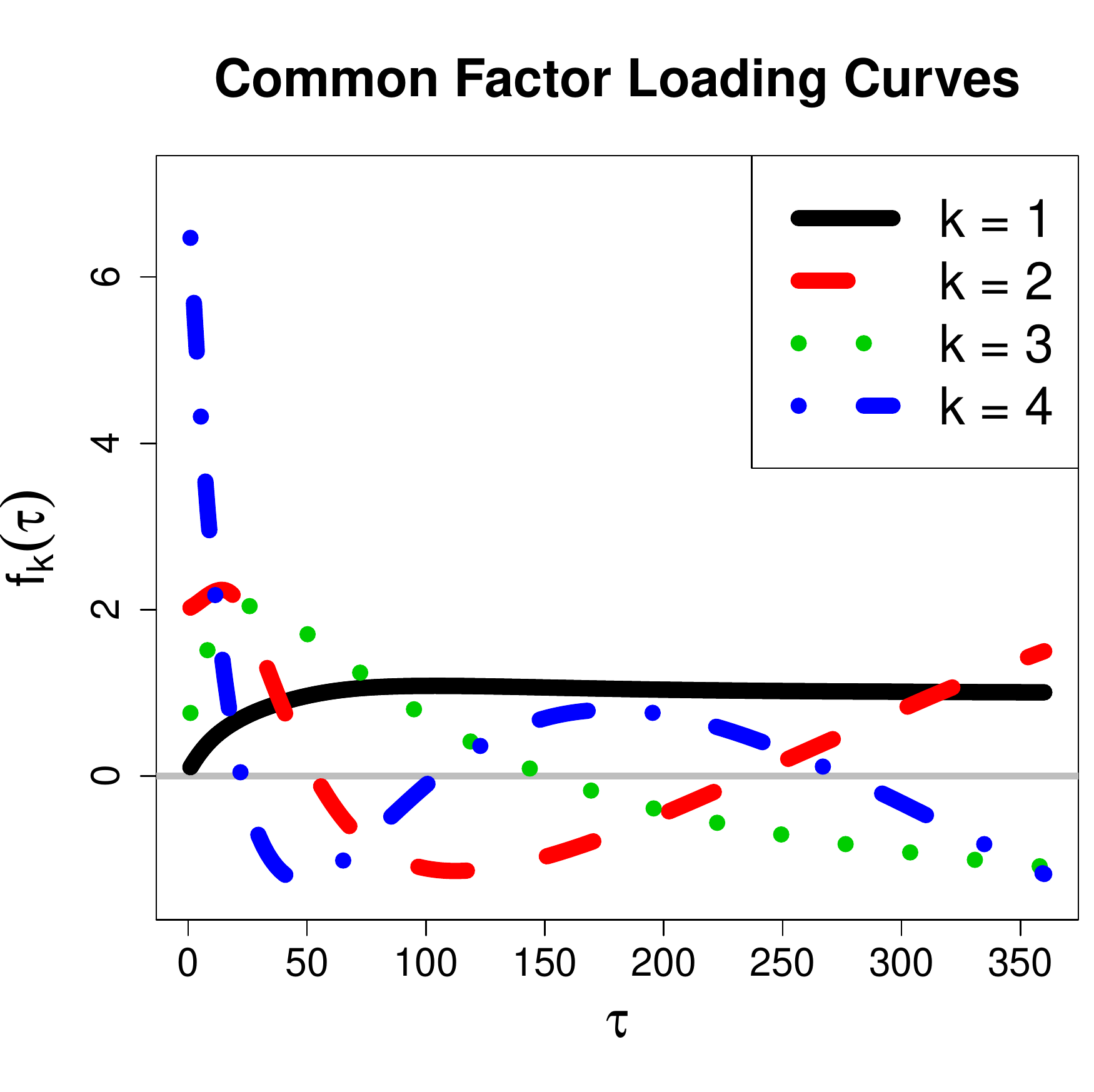} 
  \caption{Posterior means of the common FLCs, $\{f_1,f_2, f_3,f_4\}$, as a function of maturity, $\tau$.}
          \label{fig:cFLC}
\end{figure}

 \ifblinded
\color{blue} \bf
\fi

In Table 1, we  compute posterior means and 95\% highest posterior density (HPD)  intervals for $\gamma_k^{(c)}$, which measures the strength of the linear relationship between $\beta_{k,t}^{(c)}$ and $\beta_{k,t}^{(1)}$. For the level and slope factors $k=1,2$, the ECB is substantially less correlated with the Fed factors than are the BOE and BOC factors. For $k=4$, the BOE, ECB, and BOC factors are nearly uncorrelated with the Fed factors. 

\color{black}\rm

\begin{table}[ht]
\centering
\begin{tabular}{|c|cccc|}
  \hline 
  Economy  & k = 1 & k = 2 & k = 3  & k = 4 \\ 
  \hline
  \hline
  \multirow{2}{*}{BOE} & 0.62 & 0.72 & 0.37 & 0.03 \\
 & (0.57, 0.67) & (0.56, 0.89) & (0.27, 0.46) &  (-0.03, 0.09)\\ \hline 
  \multirow{2}{*}{ECB} &0.39 & 0.27 & 0.44 & 0.07 \\
 &  (0.34, 0.45) & (0.11, 0.42) & (0.35, 0.52) & (0.00, 0.15) \\ \hline 
   \multirow{2}{*}{BOC} & 0.61 & 0.56 & 0.49 & 0.16  \\
 & (0.57, 0.65)  & (0.47, 0.65) & (0.41, 0.58) & (0.08, 0.25)  \\ \hline 
\end{tabular} \caption{Posterior means and 95\% HPD intervals for $\gamma_k^{(c)}$, which measures the strength of the linear relationship between $\beta_{k,t}^{(c)}$ and $\beta_{k,t}^{(1)}$.}
\end{table}



\begin{figure}[h]
  \centering
\includegraphics[scale= .25]{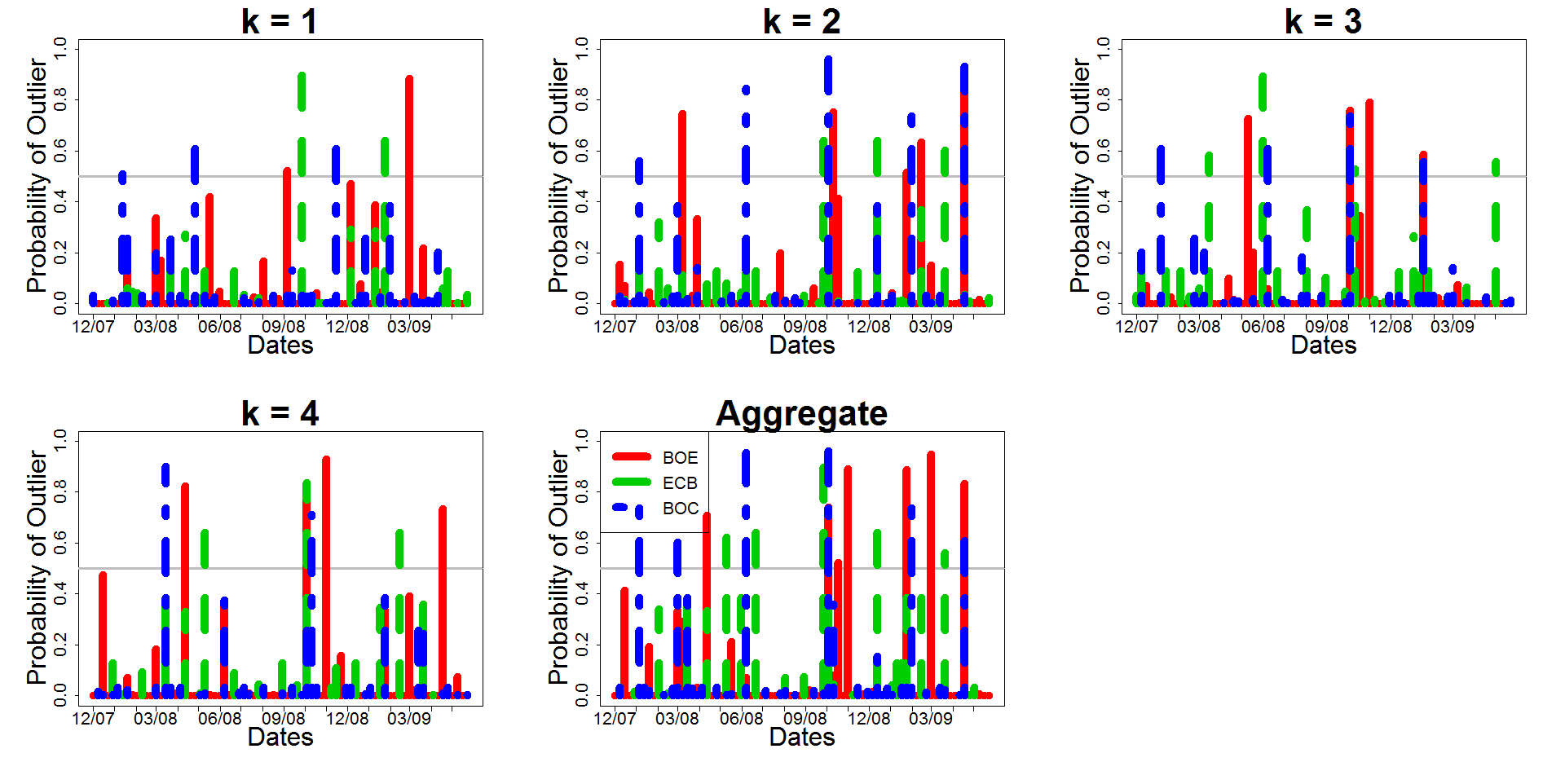} 
  \caption{The MCMC sample proportions of   $r_{k, (c),t}^2$ and $\sum_{k=1}^4 r_{k, (c), t}^2$ that exceed the  95th percentile  of the assumed $\chi^2$-distributions. }
          \label{fig:outlierPlot}
\end{figure}

 \ifblinded
\color{blue} \bf
\fi
Finally, we analyze the conditional standardized residuals from model \eqref{commonHMM}, $r_{k,(c), t} = \left(\omega_{k,t}^{(c)} - \phi_{k,1}^{(c)} \omega_{k,t-1}^{(c)}\right)/\sigma_{k, (c),t} \stackrel{iid}{\sim} N(0,1),$ to determine periods of time $t$ for which \eqref{commonHMM} is inadequate, which can indicate deviations from the assumed linear relationship between the Fed factors and the other economy factors.  
By computing the MCMC sample proportion of $r_{k,(c),t}^2\sim \chi_1^2$  that exceed a critical value of the $\chi^2$-distribution, e.g., the 95th percentile $\chi_{1, 0.05}^2 \approx 3.84$, we can obtain a simple estimate of the probability that $r_{k,(c),t}^2 $ exceeds the critical value and, by that measure, is likely an outlier. We can compute a similar quantity for $\sum_{k=1}^4 r_{k, (c), t}^2 \sim \chi_4^2$, which aggregates across factors $k=1,\ldots,4$. In Figure   \ref{fig:outlierPlot},  we plot these MCMC sample proportions, restricted to the U.S. recession of December 2007 to June 2009.  Around November 2008, there were outliers for all three economies for $k=2,3,4$  and the aggregate, which  suggests that the U.S. interest rate market may have behaved differently from the other economies during this time period. We are currently investigating an extension of model  \eqref{commonHMM} to incorporate several important financial predictors as covariates, with a particular focus on the weeks during the recession.

 \color{black}\rm

\subsection{Multivariate Time-Frequency Analysis for Local Field Potential}\label{rat}
\begin{figure}
        \centering
        \begin{subfigure}[b]{0.32\textwidth}
                \includegraphics[width=\textwidth]{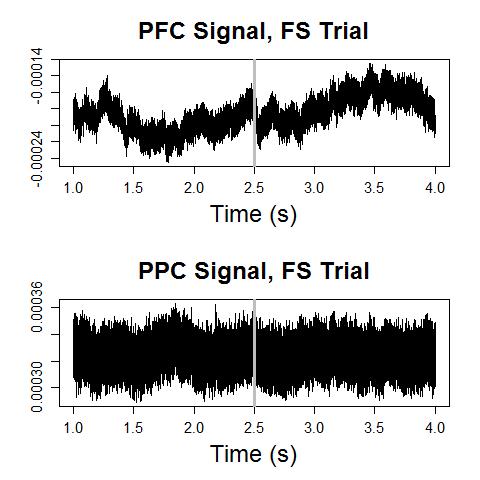}
                \caption{The bivariate LFP signal.}
                \label{fig:blfp}
        \end{subfigure}%
        ~ 
        \begin{subfigure}[b]{0.63\textwidth}
                \includegraphics[width=\textwidth]{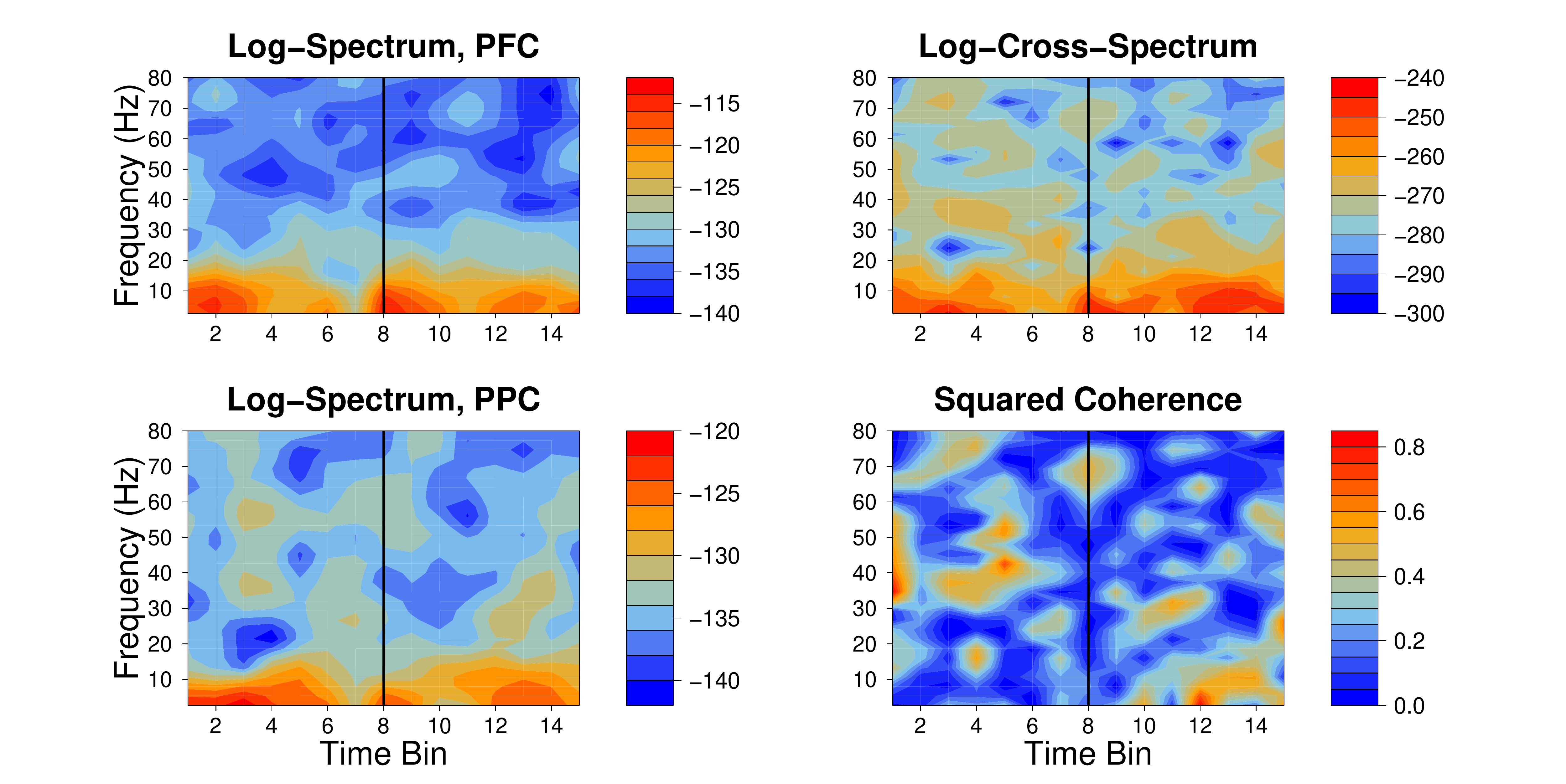}
                \caption{ The associated (log-) spectra and squared coherence.}
                \label{fig:rawspec}
        \end{subfigure}
        \caption{The raw LFP data  from a rat during an FS trial. The vertical lines indicates the approximate time at which the rat processed the stimuli, $t^*$.}
        \label{fig:raw}
\end{figure}

Local field potential (LFP) data were collected on rats to  study the neural activity involved in  {\it feature binding}, which describes how the brain amalgamates distinct sensory information into a single neural representation
(\citealp{botly2009cholinergic,vladTalk}).  
LFP uses pairs of electrodes implanted directly in local brain regions of interest to record the neural activity over time; in this case, the brain regions of interest are the prefrontal cortex (PFC) and the posterior parietal cortex (PPC). 
The rats were given two sets of tasks: one that required the rats to synthesize multiple stimuli in order to receive a reward (called {\it feature conjunction}, or FC), and one that only required the rats to process a single stimulus in order to receive a reward (called {\it feature singleton}, or FS). FC involves feature binding, while FS may serve  as a baseline. The tasks were repeated in 20 trials each for FS and FC, during which electrodes implanted in the PFC and the PPC recorded the neural activity. Therefore, the raw data signal is a bivariate time series with 40 replications for each rat; we show an example of the bivariate signals for one such replication in Figure \ref{fig:blfp}. Each signal replicate is 3 seconds long, and has been centered around the behavior-based laboratory estimate of the time at which the rat processed the stimuli, which we denote by~$t^*$.

Our interest is in the time-dependent  behavior of these bivariate signals and the interaction between them. A natural approach is to use   {\it time-frequency analysis}; however, exact inference for standard time-frequency procedures is not available. An appealing alternative is to use time-frequency methods  to transform the bivariate signal into a MFTS, which makes available the multivariate modeling and inference of the MFDLM. 

Since the MFDLM provides smoothing in both  the frequency domain $\mathcal{T}$ and  the time domain $T$, we may use time-frequency preprocessing that provides minimal smoothing.   For the time domain,  we segment the signal into time bins of width one-eighth the length of the original signal, with a 50\% overlap between neighboring bins to reduce undesirable boundary effects. 
  Within each time bin, we compute the {\it periodograms} and  {\it cross-periodogram} of the bivariate signal. Let $q_t^{(1)}(\tau)$ and $q_t^{(2)}(\tau)$ be the discrete Fourier transforms of the PFC and PPC signals, respectively, for time bin $t$ evaluated at frequency $\tau$, after removing linear trends. The periodograms are $I_t^{(c)}(\tau) =|q_t^{(c)}|^2$ for $c=1,2$ and the cross-periodogram is $I_t^{(3)}(\tau) =  q_t^{(1)}\bar{q}_t^{(2)}$, where $\bar{q}_t^{(2)}$ is the complex conjugate of ${q}_t^{(2)}$. The cross-periodogram is generally complex-valued, and if the periodograms are unsmoothed,  then $|I_t^{(3)}(\tau)|^2 = I_t^{(1)}(\tau)I_t^{(2)}(\tau)$ is real-valued but clearly fails to provide new information \citep{bloomfield2004fourier}. This does not imply that the cross-periodogram is uninformative, but rather that some frequency domain smoothing of the periodograms is necessary.

  Following \cite{shumway2000time}, we use a modified Daniell kernel to obtain the smoothed periodograms, or {\it spectra}.
 We subdivide each time bin into five segments, compute  $I_t^{(c)}(\tau), c=1,2,3$ within each segment, and then average  the resulting periodograms  using decreasing weights determined by the modified Daniell kernel. Denoting these spectra by $\tilde{I}_t^{(c)}(\tau)$, we let $Y_t^{(c)}(\tau) = \log\left(\tilde{I}_t^{(c)}(\tau)\right)$ for $c=1,2$, where the log-transformation is appealing because it is the variance-stabilizing transformation for the periodogram \citep{shumway2000time}. To account for the periodic dependence between signals, one choice is the log-cross-spectrum, $\log\left(|\tilde{I}_t^{(3)}(\tau)|^2\right)$. An appealing alternative is  the {\it squared coherence} defined by $\kappa_{t}^2(\tau) \equiv |\tilde{I}_t^{(3)}(\tau)|^2/ (\tilde{I}_t^{(1)}(\tau)\tilde{I}_t^{(2)}(\tau))$, which satisfies the constraints $0\le \kappa_{t}^2(\tau) \le 1$  and is the frequency domain analog to the squared correlation \citep{bloomfield2004fourier}.  Since \eqref{fhdlm} specifies that $Y_t^{(c)}(\tau) \in \mathbb{R}$, we transform the squared coherence and let $Y_t^{(3)}(\tau) = \Phi^{-1}(\kappa_t^2(\tau)) \in \mathbb{R}$, where $\Phi^{-1}:[0,1]\rightarrow \mathbb{R}$ is a known monotone function;  we use the Gaussian quantile function. We have found that fitting $Y_t^{(3)}(\tau)$ produces very similar results to fitting $\kappa_t^2(\tau)$ directly, yet in the transformed case, our estimate of the squared coherence $\Phi\left(\mu_t^{(3)}(\tau)\right)$ obeys the constraints. Because of our Bayesian approach, this transformation does not inhibit inference.
  
 More generally, this procedure is applicable to $\ell$-dimensional time series, which, including either the squared coherence or the cross-spectra, yields a $C= \ell(\ell+1)/2$-dimensional MFTS.  We show an example of the resulting MFTS from a rat during an FS trial in Figure \ref{fig:rawspec}. For completeness, we  include the log-cross-spectrum, which is not a component of the MFTS.

\subsubsection{MFDLM Specification} \label{ratmodsec}
We use the common FLCs model of Section \ref{comcurve}  accompanied by a random walk model for  the factors:
\begin{equation}\label{ratmod}
\begin{cases}
Y_{i,s,t}^{(c)}(\tau) = \sum_{k=1}^K \beta_{k,i,s,t}^{(c)} f_k(\tau) + \epsilon_{i,s,t}^{(c)}(\tau),  & \left[\epsilon_{i,s,t}^{(c)}(\tau)\big| \sigma_{(c)}^2\right] \stackrel{indep}{\sim} N(0, \sigma_{(c)}^2) \\
\utwi{\beta}_{k,i,s,t} =   \utwi{\beta}_{k,i,s,t-1} + \utwi{\omega}_{k,i,s,t}, & \left[\utwi{\omega}_{k,i,s,t}\big|  \mathbf{W}_k\right] \stackrel{indep}{\sim} N(\mathbf{0}, \mathbf{W}_k) 
\end{cases}
\end{equation}
where $\utwi{\beta}_{k,i,s,t} = (\beta_{k,i,s,t}^{(1)}, \ldots, \beta_{k,i,s,t}^{(C)})'$, $Y_{i,s,t}^{(c)}$  are the  log-spectra for $c=1,2$ and    the probit-transformed squared coherences for $c=3$, $i =1 ,\ldots,8$ index the rats, $s=1,\ldots,40$ index the trials for each rat, and $t=1,\ldots, 15$ index the time bins for each trial. The joint indices $(i,s,t)$ in \eqref{ratmod} correspond to the time index $t$ in \eqref{fhdlm}, and are used to specify independence of the residuals $\utwi{\omega}_{k,i,s,t}$ between rats and between trials. 
 \ifblinded
\color{blue} \bf
\fi
For each initial time bin $t=1$, we let  $\utwi{\beta}_{k,i,s,1} \sim N(\utwi{0}, 10^4 \mathbf{I}_{C\times C})$, since the corresponding observations   are only time-ordered    {\it within} a trial. 
\color{black}\rm
The $C\times C$ factor covariance matrices $\mathbf{W}_k$ do not depend on the rat or the trial, and  can help summarize the overall dependence among factors. For simplicity and parsimonious modeling, \eqref{ratmod} assumes independence between $\utwi{\omega}_{k,i,s,t}$ and $\utwi{\omega}_{j,i,s,t}$ for $j\ne k \in \{1,\ldots,K\}$, but allows for correlation between outcomes for fixed $k$.  The $\mathbf{W}_k$ control the amount of time domain smoothing  for the factors and therefore for $\mu_{i,s,t}^{(c)}(\tau) \equiv \sum_{k=1}^K \beta_{k,i,s,t}^{(c)} f_k(\tau)$. 
 \ifblinded
\color{blue} \bf
\fi
For the error variances, we use the   conjugate priors $\sigma_{(c)}^{-2} \stackrel{iid}{\sim} \mbox{Gamma}\left(0.001, 0.001\right)$ and $\mathbf{W}_k^{-1} \stackrel{iid}{\sim} \mbox{Wishart} ((\rho R)^{-1}, \rho)$, with $R^{-1} = \mathbf{I}_{C\times C}$,    the expected prior precision,  and $\rho = C \ge \mbox{rank}(R^{-1})$. We provide the full conditional posterior distributions  in the appendix.

\color{black}\rm


To determine the effects of feature binding, we compare the values of  $\mu_{i,s,t}^{(c)}(\tau)$ between the FS and FC trials. Letting $S_{i,FC}$ (respectively, $S_{i,FS}$) be the subset of FC (respectively, FS) trials for which rat $i$ received the reward, we estimate posterior distributions for the sample means $\bar{\mu}_{t}^{(c)}(\tau) \equiv \frac{1}{8}\sum_{i=1}^{8} \left[ \frac{1}{|S_{i,FC}|} \sum_{s \in S_{i,FC}} \mu_{i,s,t}^{(c)}(\tau) - \frac{1}{|S_{i,FS}|} \sum_{s' \in S_{i,FS}}  \mu_{i,s',t}^{(c)}(\tau) \right]$ for $c=1,2$ and $\bar{\mu}_{t}^{(3)}(\tau) \equiv \frac{1}{8}\sum_{i=1}^{8} \left[ \frac{1}{|S_{i,FC}|} \sum_{s \in S_{i,FC}} \Phi\left(\mu_{i,s,t}^{(3)}(\tau)\right) - \frac{1}{|S_{i,FS}|} \sum_{s' \in S_{i,FS}} \Phi\left(\mu_{i,s',t}^{(3)}(\tau)\right) \right]$. Therefore, we examine the difference in the log-spectra and the squared coherences between the FC and the FS trials, which we average over all rats and over all trials for which the rat responded {\it correctly} to the stimuli. This restriction is important, since it filters out unrepresentative trials, in particular  FC trials for which feature binding may not have occurred.  

\subsubsection{Results}
Since we observe functions in 15 time bins for 40 trials for 8 rats, the time-dimension of our 3-dimensional MFTS is $T=(15)(40)(8) = 4800$. We restrict the frequencies to $\mathcal{T} = [0.1, 80]$ Hz, which is the range of interest for this application and yields $m_t^{(c)} =   30$ for all $c,t$. Guided by DIC, we select $K=10$. Alternatively, we could use a smaller value of $K$ by increasing the initial smoothing of the log-spectra and the  squared coherences, but would risk smoothing over important features. 
 \ifblinded
\color{blue} \bf
\fi
We ran the MCMC sampler for $7,000$ iterations and discarded the first $2,000$ iterations as a burn-in; see the appendix for the MCMC diagnostics. 

\color{black}\rm

We compute 95\% pointwise HPD intervals and posterior means for $\bar{\mu}_t^{(c)}(\tau)$, $c=1,2,3$ and display the results as spectrogram plots; the plots for $c=1,2$ are in the appendix, while $c=3$ is in Figure \ref{fig:coh}. Regions of red or orange in the lower 95\% HPD interval plots indicate a significant positive difference between the FC and FS trials, while regions of blue in the upper 95\% HPD interval plots indicate a significant negative difference. We are particularly interested in the  time bins around $t^*$, which indicates the approximate time at which the stimuli were processed, and frequencies up to 40-50 Hz.

The averages of the differenced log-spectra,  $\bar{\mu}_t^{(1)}(\tau)$ and  $\bar{\mu}_t^{(2)}(\tau)$, describe how the distinct regions of the brain|the PFC and PPC, respectively|respond differently to stimuli that do or do  not require feature binding. 
By comparison, the average of the differenced squared coherences,  $\bar{\mu}_t^{(3)}(\tau)$, describes how these regions of the brain interact with each other under the different stimuli. Based on Figure \ref{fig:coh},   feature binding appears to be most strongly associated with greater squared coherence at frequencies in the Theta range (4-8 Hz), the Alpha range (8-13 Hz), and the Beta range (13-30 Hz)  around $t^*$. This pattern persists in the power of both the PFC and PPC log-spectra plots, which suggests that these ranges of frequencies are important to the process of feature binding. Therefore, using the inference provided by the MFDLM, we conclude that during feature binding, the Theta, Alpha, and Beta ranges  are associated with increased brain activity in both the PFC and the PPC, as well as  greater synchronization between these regions.

  \begin{figure}[h] 
  \centering
    \includegraphics[scale= .3]{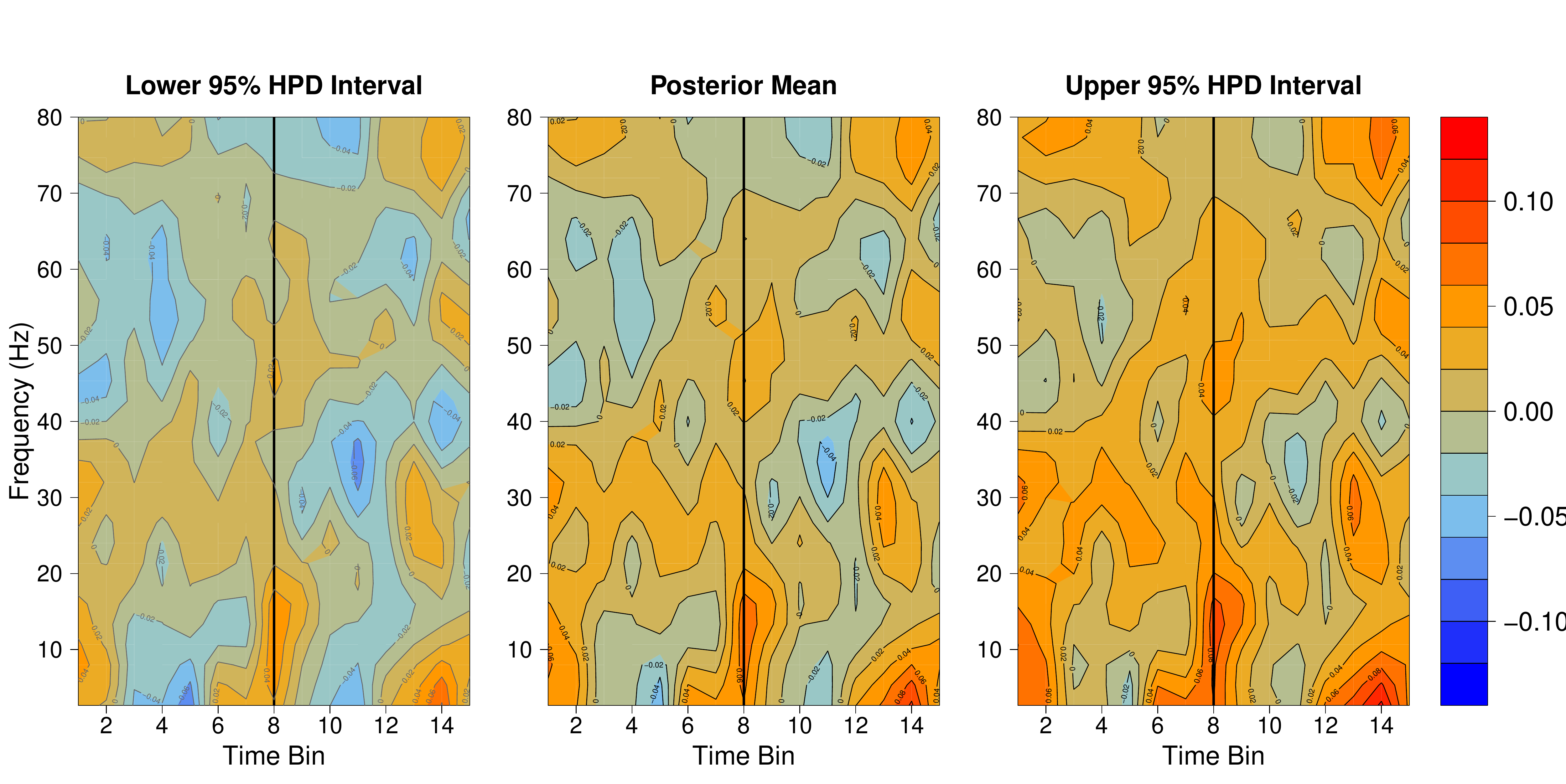} 
  \caption{Pointwise 95\% HPD intervals and the posterior mean for $\bar{\mu}_t^{(3)}$, which is the average difference in squared coherence between the FC and FS trials. The black vertical lines indicate the event time  $t^*$.}
          \label{fig:coh}
\end{figure}

\section{Conclusions}
The MFDLM provides a general framework to model complex dependence  among functional observations. Because we separate out the functional component through appropriate conditioning and include the necessary identifiability constraints, we can model the remaining dependence using familiar scalar and multivariate methods. The hierarchical Bayesian approach allows us to  incorporate interesting and useful submodels seamlessly, such as the common trend  model of Section \ref{comHMM}, the stochastic volatility model of Section \ref{svm}, and the random walk model of Section \ref{ratmodsec}. We combine Bayesian spline theory and convex optimization  to model the functional component as a set of smooth and optimal curves subject to (identifiability) constraints. Using an efficient Gibbs sampler, we obtain posterior samples of all of the unknown parameters in \eqref{fhdlm}, which allows us to perform 
 inference on any  parameters of interest, such as $\bar{\mu}_t^{(c)}$ in the LFP example.

Our two diverse applications demonstrate the flexibility and wide applicability of our model. The common trend  model of Section \ref{comHMM} provides useful insights into the interactions among multi-economy yield curves, and our LFP example suggests a novel approach to time-frequency analysis via MFTS. In these applications, the MFDLM adequately models a variety of functional dependence structures, including time dependence, (time-varying) contemporaneous dependence, and stochastic volatility, and may readily accommodate additional dependence structures, such as covariates, repeated measurements, and spatial dependence. We are currently developing an \texttt{R} package for our methods.

\bibliographystyle{apalike}
\bibliography{BayesianMFDLMbib}

\appendix

\section{Appendix}

To sample from the joint posterior distribution, we use a Gibbs sampler. Because the Gibbs sampler allows blocks of parameters to be conditioned on all other blocks of parameters, it is a convenient approach for our model. First, hierarchical dynamic linear model (DLM) algorithms typically require that $\utwi{\beta}_t$ and $\utwi{\theta}_t$ be the only unknown components, which we can accommodate by conditioning appropriately. Second, our orthonormality approach for $f_k^{(c)}$ fits nicely within a Gibbs sampler, and we can adapt the algorithms described in 
\cite{wand2008semiparametric}. 
And third, the hierarchical structure of our model imposes natural conditional independence assumptions, which allows us to easily partition the parameters into appropriate blocks. 

\numberwithin{equation}{subsection}
\numberwithin{table}{subsubsection}
\numberwithin{figure}{subsubsection}

\subsection{Initialization}\label{inits}
To initialize  the factors $\utwi{\beta}_k^{(c)} = \left(\beta_{k,1}^{(c)}, \ldots,\beta_{k,T}^{(c)}\right)'$ and the factor loading curves (FLCs) $f_k^{(c)}$ for $k=1,\ldots,K$ and $c=1,\ldots,C$, we compute the singular value decomposition (SVD) of the data matrix $\mathbf{Y}^{(c)} = \mathbf{U}^{(c)} \utwi{\Sigma}^{(c)} {\mathbf{V}^{(c)}}'$ for  $c=1,\ldots,C$. Note that to obtain a data {\it matrix} $\mathbf{Y}^{(c)}$, with rows corresponding to times $t$ and columns to observations points $\tau$, we need to estimate  $Y_t^{(c)}(\tau)$ for any unobserved $\tau$ at each time $t$, which may be computed quickly using splines. However, these estimated data values are {\it only} used for the initialization step. Letting $\mathbf{U}_{1:K}^{(c)}$ be the first $K$ columns of $\mathbf{U}^{(c)}$,  $\utwi{\Sigma}_{1:K}^{(c)}$ be the upper left $K\times K$ submatrix of $\utwi{\Sigma}^{(c)}$, and $\mathbf{V}_{1:K}^{(c)}$   be the first $K$ columns of $\mathbf{V}^{(c)}$, we initialize the factors $\left(\utwi{\beta}_1^{(c)},\ldots,\utwi{\beta}_K^{(c)}\right) = \mathbf{U}_{1:K}^{(c)}\utwi{\Sigma}_{1:K}^{(c)}$ and the FLCs $\left(\utwi{f}_1^{(c)},\ldots,\utwi{f}_K^{(c)}\right) = \mathbf{V}_{1:K}^{(c)}$, where  $\utwi{f}_k^{(c)}$ is the vector of FLC $k$ evaluated at all  observation points $\cup_t \mathcal{T}_t^{(c)}$ for outcome $c$. The $\utwi{f}_k^{(c)}$ are orthonormal in the sense that ${\utwi{f}_k^{(c)}}'\utwi{f}_j^{(c)} = \mathbf{1}\!(k=j)$, but they are not smooth. This approach is similar to the initializations in \cite{matteson2011forecasting} and \cite{FDFM}.

Given the factors $\utwi{\beta}_k^{(c)}$ and the FLCs $\utwi{f}_k^{(c)}$, we can estimate each $\sigma_{(c)}^2$ (or more generally, $\mathbf{E}_t$) using conditional maximum likelihood, with the likelihood from the observation level of model \eqref{fhdlm}. Similarly, we can estimate each $\lambda_{k, (c)}$ conditional on $\utwi{f}_k^{(c)}$ by maximizing the   likelihood $\utwi{d}_k^{(c)} \sim N(\utwi{0},\mathbf{D}_k^{(c)})$ with respect to $\lambda_{k, (c)}$, where $\mathbf{D}_k^{(c)} = \mbox{diag}\left(10^8, 10^8, \lambda_{k,(c)}^{-1}, \ldots,  \lambda_{k,(c)}^{-1}\right)$. Then, given $\lambda_{k, (c)}$, $\sigma_{(c)}^2$,  $\utwi{\beta}_k^{(c)}$, and $\utwi{f}_k^{(c)}$, we can estimate each $\utwi{d}_k^{(c)}$  by normalizing the full conditional posterior expectation given in the main paper; i.e., solving the relevant quadratic program and then normalizing the solution. 
Initializations for the remaining levels proceed similarly as conditional MLEs, but depend on the form chosen for $\mathbf{X}_t$, $\mathbf{V}_t$, $\mathbf{G}_t$,  and $\mathbf{W}_t$. In our applications, this  conditional MLE approach produces  reasonable starting values for all variables.

\subsubsection{Common Factor Loading Curves}
If we wish to implement the common FLCs model $f_k^{(c)} = f_k$ for all $k,c$, then we instead compute the SVD of the stacked data matrices $ \left({\mathbf{Y}^{(1)}}', \ldots, {\mathbf{Y}^{(C)}}'\right)' = \mathbf{U}\utwi{\Sigma}\mathbf{V}'$, where now the data matrices $\mathbf{Y}^{(1)}, \ldots, \mathbf{Y}^{(C)}$ are imputed using splines for all observation points for all outcomes,  $\cup_{t,c} \mathcal{T}_t^{(c)}$, and therefore have the same number of columns. Alternatively, we may improve computational efficiency by choosing a small yet representative subset of observation points $\mathcal{T}^* \subset \cup_{t,c} \mathcal{T}_t^{(c)}$ and then estimating each data matrix $\mathbf{Y}^{(c)}$ for all   $\tau \in \mathcal{T}^*$. Let $\mathbf{U}_{1:K}^{(c)}$ be the first $K$ columns of $\mathbf{U}^{(c)}$,  where $\mathbf{U}^{(c)}, c=1,\ldots,C,$ corresponds to the outcome-specific blocks of $\mathbf{U} = \left({\mathbf{U}^{(1)}}',\ldots,{\mathbf{U}^{(C)}}'\right)'$. Then, similar to before, we set  $\left(\utwi{\beta}_1^{(c)},\ldots,\utwi{\beta}_K^{(c)}\right) = \mathbf{U}_{1:K}^{(c)}\utwi{\Sigma}_{1:K}$ for $c=1,\ldots,C$, and $(\utwi{f}_1,\ldots,\utwi{f}_K) = \mathbf{V}_{1:K}$, where  $\utwi{\Sigma}_{1:K}$ is the upper left $K\times K$ submatrix of $\utwi{\Sigma}$ and  $\mathbf{V}_{1:K}$   is the first $K$ columns of $\mathbf{V}$. Again, the  $\utwi{f}_k$ are unsmoothed with  $\utwi{f}_k'\utwi{f}_j = \mathbf{1}\!(k=j)$, but now the initialized FLCs are common for $c=1,\ldots,C$. Initialization of the remaining parameters proceeds as before, but now with $\lambda_{k,(c)} = \lambda_k$ and $\utwi{d}_k^{(c)} = \utwi{d}_k$, which can be obtained by maximizing the relevant conditional likelihoods under the common FLCs model. 

\subsubsection{Computing a range for $K$}
The initialization procedure requires the SVD of the data matrix.
If we first center the columns of the data matrix, then the squared components of the diagonal matrix $\utwi{\Sigma}^{(c)}$ (or $\utwi{\Sigma}$) indicate the variance explained by each factor. Therefore, we can estimate the proportion of total variance in the data explained by each factor, without the need to run an MCMC sampler. Using this information, we can either select $K$ based on the minimum number of factors needed to explain a prespecified proportion of total variance explained, such as $95\%$, or select a range for $K$ based on an interval of proportion of total variance explained, such as $(80\%, 99\%)$. In the latter case, we can then select $K$ by comparing the marginal likelihood or DIC for each $K$ in this range. Note that in both cases, it may be appropriate to increase the selected value(s) of $K$ by one to account for the initial centering of the data matrix.

\subsection{Sampling the MFDLM} \label{algorithm}
For greater generality, we present our sampling algorithm for  non-common FLCs; i.e., we retain dependence on $c$ for $\utwi{d}_k^{(c)}$ and $\lambda_{k,(c)}$. When applicable, we discuss the necessary modifications for the common FLCs model. 

The algorithm proceeds in four main blocks:

\begin{enumerate}
\item Sample the smoothing parameters $\lambda_{k,(c)}$ and the basis coefficients $\utwi{d}_k^{(c)}$     for the FLCs. Using uniform priors on the standard deviations $\lambda_{k,(c)}^{-1/2}$ and enforcing the ordering constraints $\lambda_{1,(c)} > \lambda_{2,(c)} > \cdots > \lambda_{K,(c)}$, the conditional priors are $\lambda_{k,(c)}^{-1/2} \sim \mbox{Uniform}\left(\ell_{k, (c)}, u_{k,(c)}\right)$, where $\ell_{1, (c)}=0$, $\ell_{k, (c)} = \lambda_{k-1, (c)}^{-1/2}$ for $k=2,\ldots,K$, $u_{k, (c)} =  \lambda_{k+1, (c)}^{-1/2}$ for $k=1,\ldots, K-1$, and $u_{K, (c)} = 10^4$. For $k=1,\ldots,K, c=1,\ldots,C$, the full conditional distribution  for $\lambda_{k, (c)}$ is  $
 \mbox{Gamma}\left( \frac{1}{2}(M+1), \frac{1}{2} \sum_{j=3}^{M+4} d_{k,(c), j}^2\right)$
  truncated to the interval $(u_{k, (c)}^{-2}, \ell_{k, (c)}^{-2})$, where $M$ is the number of interior knots, $d_{k,(c), j}$ are the components of $\utwi{d}_k^{(c)}$, and $\ell_{1,(c)}^{-2} = \infty$.  For the common FLCs model, we simply replace $\utwi{d}_k^{(c)}$ with $\utwi{d}_k$ to obtain the full conditional posterior for $\lambda_k$.    To reduce dependence of the ordering of   $\lambda_{k,(c)}$ on the initialization procedure of Section \ref{inits}|which fixes the ordering without accounting   for the smoothness of the FLCs $f_k^{(c)}$|we run the first 10 MCMC iterations without enforcing the ordering constraints, so $\ell_{k, (c)}=0 $ and $u_{k,(c)} = 10^4$ for $k=1,\ldots,K$.  At the end of this brief trial run, we reorder $\lambda_{k, (c)}, f_k^{(c)}$, and $\beta_{k,t}^{(c)}$ to reflect the ordering constraint; we may reorder the other parameters as well, but typically this is not necessary.   
  We can sample $\lambda_{k,(c)}$ from the truncated Gamma distribution using the following procedure:
\begin{enumerate}
\item Sample $U \sim \mbox{Uniform}\left(a, b\right)$, where $a = F_G(u_{k,(c)}^{-2}) $ and $b = F_G(\ell_{k,(c)}^{-2}) $, with $F_G(\cdot)$ the distribution function of the full conditional Gamma distribution given above;
\item Set $\lambda_{k, (c)} = F_G^{-1}(U)$.
\end{enumerate}

After sampling the $\lambda_{k,(c)}$, we sample and then normalize the  $\utwi{d}_k^{(c)}$ with a modified version of the efficient Cholesky decomposition approach of \cite{wand2008semiparametric}:
\begin{enumerate}
\item Compute the (lower triangular) Cholesky decomposition $\mathbf{B}_k^{-1}   = \bar{\mathbf{B}}_L\bar{\mathbf{B}}_L'$;
\item Use forward substitution to obtain $\bar{\mathbf{b}}$ as the solution to $\bar{\mathbf{B}}_L \bar{\mathbf{b}} = \mathbf{b}_k$,  then use backward substitution to obtain $\utwi{d}_k^U$ as the solution to $\bar{\mathbf{B}}_L'\utwi{d}_k^U = \bar{\mathbf{b}} + \bar{\mathbf{z}}$, where $\bar{\mathbf{z}}\sim N (\mathbf{0},  \mathbf{I}_{ (M+4)\times (M+4)})$;
\item Use forward substitution to obtain $\bar{\mathbf{L}}$ as the solution to $\bar{\mathbf{B}}_L \bar{\mathbf{L}}= \mathbf{L}_{[-k]} $,  then use backward substitution to obtain $\mathbf{\tilde{L}}$ as  the solution to $\bar{\mathbf{B}}_L'\mathbf{\tilde{L}} = \bar{\mathbf{L}}$;
\item Set $\utwi{d}_k^* = \utwi{d}_k^U - \mathbf{\tilde{L}} (\mathbf{L}_{[-k]}'\mathbf{\tilde{L}})^{-1} \mathbf{L}_{[-k]}' \utwi{d}_k^U$;
\item Retain the vector $\utwi{d}_k^{(c)}=\utwi{d}_k^* / \sqrt{{\utwi{d}_k^*}' \mathbf{J}_\phi \utwi{d}_k^*}$ and set $\utwi{\beta}_k^{(c)} = \sqrt{{\utwi{d}_k^*}' \mathbf{J}_\phi \utwi{d}_k^*} \utwi{\beta}_k^{(c)}$.
\end{enumerate} 
The definitions of $\mathbf{B}_k$ and $\mathbf{b}_k$ depend on whether or not we use the common FLCs model with $f_k^{(c)} = f_k$ (see Section 3 of the paper). 
The sample $\utwi{d}_k^U \sim N(\mathbf{B}_k\mathbf{b}_k, \mathbf{B}_k)$ in (b) is unconstrained, while steps (c) and (d) incorporate the linear orthogonality constraints:  the random variable $\utwi{d}_k^* = \utwi{d}_k^U - \mathbf{B}_k \mathbf{L}_{[-k]}(\mathbf{L}_{[-k]}'\mathbf{B}_k \mathbf{L}_{[-k]})^{-1}\mathbf{L}_{[-k]}'\utwi{d}_k^U$ follows the correct distribution $N(\mathbf{\tilde{B}}_k, \mathbf{b}_k, \mathbf{\tilde{B}}_k)$, which conditions on the   linear orthogonality constraints $\utwi{d}_k' \mathbf{L}_{[-k]} = \utwi{0}$. Steps (c) and (d) compute this random variable efficiently  (see \citealp[Chapter~12]{gelfand2010handbook}   for more details).
The scaling of $\utwi{d}_k^{(c)}$ and $\utwi{\beta}_k^{(c)}$ in (d) enforces the unit-norm constraint on $f_k^{(c)}$ yet ensures that  $f_k^{(c)}(\tau) \utwi{\beta}_k^{(c)}$|which appears in the posterior distribution of $\utwi{d}_{j}^{(c)}$ for all $j\ne k$|is unaffected by the normalization. To encourage better mixing, we randomly select the order of $k=1,\ldots,K$ in which to sample $\lambda_{k, (c)} $ and $ \utwi{d}_k^{(c)}, c=1,\ldots,C$.

\item Sample  the factors $\utwi{\beta}_t$ (and  $\utwi{\theta}_t$, if present) conditional on all other parameters in \eqref{fhdlm} using  the state space sampler of \cite{durbin2002simple,koopman2003filtering,koopman2000fast},   which is optimized when $\mathbf{E}_t$ is diagonal. For general hierarchical models, we may modify the hierarchical DLM algorithms of \cite{gamerman1993dynamic}. 

For the prior distributions, we only need to specify the distribution of $\utwi{\beta}_0$ (and $\utwi{\theta}_0$); the remaining distributions are computed recursively using  $\mathbf{F}$, $\mathbf{X}_t$, $\mathbf{G}_t$ and the error variances. For simplicity, we let $\beta_{k,0}^{(c)} \stackrel{iid}{\sim} N(0, 10^4)$, which is a common  choice for DLMs.  

\item Sample the state evolution matrix $\mathbf{G}_t$ (if unknown). $\mathbf{G}_t$ may have a special form (see Section \ref{commontrend} of this supplement) or provide a more common time series model such as a VAR. In the latter case, we may choose some structure for $\mathbf{G}_t = \mathbf{G}$, e.g. diagonality to allow dependence between $\beta_{k,t}^{(c)}$ and $\beta_{k,t-1}^{(c)}$, or $K$ blocks of dimension $C\times C$  to  allow dependence between $\beta_{k,t}^{(c)}$ and $\beta_{k,t-1}^{(c')}$ for $c,c' =1,\ldots,C $. It is particularly convenient to assume a Gaussian prior for the  nonzero entries of $\mathbf{G}$, which is a conjugate prior for $\mbox{vec}_0\left(\mathbf{G}\right)$, where $\mbox{vec}_0$ stacks the nonzero entries of the matrix (by column) into a vector.

\item  Sample each of the remaining error variance parameters separately: $\mathbf{E}_t$, $\mathbf{V}_t$, and $\mathbf{W}_t$. These distributions depend on our assumptions for the model structure, but we typically prefer conjugate priors when available. In both applications, we fix $\mathbf{V}_t = \utwi{0}_{CK\times CK}$ to remove a level in the hierarchy, and let $\mathbf{E}_t  = \mbox{diag}\left(\sigma_{(1)}^2, \ldots, \sigma_{(C)}^2\right)$ with $\sigma_{(c)}^{-2} \stackrel{iid}{\sim} \mbox{Gamma}\left(0.001, 0.001\right)$, for which the full conditional posterior distribution is $$\mbox{Gamma}\left(0.001 + \frac{1}{2}\sum_{t \in T^{(c)}} \left|\mathcal{T}_t^{(c)}\right|, 0.001 + \frac{1}{2} \sum_{t \in T^{(c)}} \sum_{\tau \in \mathcal{T}_t^{(c)}} \left\{Y_t^{(c)}(\tau) - \sum_{k=1}^K \beta_{k,t}^{(c)}f_k^{(c)}(\tau)\right\}^2\right).$$ In the random walk factor model of \eqref{ratmod}, we have $\utwi{\beta}_{k,i,s,t} =   \utwi{\beta}_{k,i,s,t-1} + \utwi{\omega}_{k,i,s,t}$ with $ \utwi{\omega}_{k,i,s,t} \stackrel{indep}{\sim} N(\mathbf{0}, \mathbf{W}_k)$ for $t=2,\ldots,15$. Using the Wishart prior $\mathbf{W}_k^{-1} \sim \mbox{Wishart} ((\rho R)^{-1}, \rho)$,  the full conditional posterior distribution for the precision is
$\mathbf{W}_k^{-1} \sim \mbox{Wishart} ((\rho R + \sum_{i,s,t} \mathbf{w}_{k,i,s,t}\mathbf{w}_{k,i,s,t}')^{-1}, \rho + 4480)$, where $\mathbf{w}_{k,i,s,t} = \utwi{\beta}_{k,i,s,t} - \utwi{\beta}_{k,i,s,t-1}$ is conditional on the factors and 4480 counts the indices $(i,s,t)$ in the summation.
We let $R^{-1} = \mathbf{I}_{C\times C}$, which is the expected prior precision, and $\rho = C \ge \mbox{rank}(R^{-1})$. 

For the  stochastic volatility model of Section \ref{svm}, we use the prior distributions and sampling algorithm given in \cite{kastner2014ancillarity}, implemented via the \texttt{R} package \texttt{stochvol} \citep{kastner2015stochvol}. Letting $\sigma_{k,(c),t}^2 = \exp(h_{k,t}^{(c)})$,   the  model is
$ h_{k,t}^{(c)} = \xi_{k,0}^{(c)} + \xi_{k,1}^{(c)}(h_{k,t-1}^{(c)} - \xi_{k,0}^{(c)}) +  \zeta_{k,t}^{(c)}$, where $\zeta_{k,t}^{(c)} \stackrel{indep}{\sim} N(0, \sigma_{H,k,(c)}^2)$ for $t=2,\ldots,T$ and $ h_{k,1}^{(c)} \sim N\!( \xi_{k,0}^{(c)}, \sigma_{H, k,(c)}^2/(1-(\xi_{k,1}^{(c)})^2))$ with $|\xi_{k,1}^{(c)}| < 1 $ for stationarity. The accompanying priors are $\xi_{k,0}^{(c)} \stackrel{indep}{\sim} N(0, 10^4)$, $(\xi_{k,1}^{(c)} + 1)/2 \stackrel{indep}{\sim} \mbox{Beta}\left(5, 1.5\right)$, and $\sigma_{H, k,(c)}^2 \stackrel{indep}{\sim} \mbox{Gamma}\left(\frac{1}{2}, \frac{1}{2}\right).$ The hyperparameters for the Beta prior are chosen reflect the high persistence of volatility commonly found in financial data, and the prior for $\sigma_{H, k,(c)}^2$ corresponds to a half-normal distribution.
For additional motivation for the stochastic volatility approach over GARCH models, see \cite{dan1998}. Note that the sampling algorithm of \cite{kastner2014ancillarity} requires a Metropolis step, and therefore the methods of  \cite{chib2001marginal} are more appropriate for marginal likelihood computations. 

\end{enumerate}

Recall that we construct a posterior distribution of $\utwi{d}_k^{(c)}$  without  the unit norm constraint, and then normalize the samples from this distribution. As a result, the conditions of Theorem 1 are satisfied and the (unnormalized) full conditional posterior distribution of $\utwi{d}_k^{(c)}$ is Gaussian, both of which are convenient results. The normalization step 1.(d) is interpretable, corresponding to the projection of a Gaussian distribution onto the unit sphere. Note that rescaling the factors  $\utwi{\beta}_k^{(c)}$ in 1.(d) does not affect the remainder of the sampling algorithm (steps 2. - 4.). The rescaled $\utwi{\beta}_k^{(c)}$ are from the previous MCMC iteration, which does not affect the  full conditional distributions of step 2. in the current MCMC iteration. The subsequent steps 3., 4., and 1. are then conditional on the newly sampled factors $\utwi{\beta}_k^{(c)}$ from step 2., which have not been rescaled.


\subsection{MCMC Diagnostics}
To demonstrate convergence and efficiency of the Gibbs sampler, we provide MCMC diagnostics for both applications. We include trace plots for several variables of interest to asses the mixing and convergence of the simulated chains. The trace plots also suggest reasonable lengths of the burn-in, i.e., the initial simulations that are discarded prior to convergence of the chain. To measure the efficiency of the sampler, we compute  the ratio of the effective sample size to the simulation sample size for several variables. We refer to this quantity as the efficiency factor,  which is the reciprocal of the simulation inefficiency factor (e.g., \citealp{kim1998stochastic}). 
All diagnostics were computed using the  \texttt{R} package  \texttt{coda}    \citep{plummer2006coda}.

\subsubsection{Multi-Economy Yield Curves}
We ran the MCMC sampler for  7,000  iterations and discarded the first 2,000 iterations as a burn-in. Longer chains and dispersed starting values did not produce noticeably different results.  The sampler was run in \texttt{R}, and took 181 minutes on a laptop with a 2.40 GHz Intel i7-4700MQ CPU using one core.  We are currently developing an \texttt{R} package for the MFDLM sampler, and expect sizable gains in computational efficiency by coding the algorithms in \texttt{C}. 

Tables A.3.1.1, A.3.1.2, and A.3.1.3 contain the efficiency factors for the common  FLCs $f_k$ evaluated at several quantiles of $\tau$, the factors $\beta_{k,t}^{(c)}$ at  various times $t$,    and the slopes $\gamma_k^{(c)}$ from the common trend model, respectively.  The efficiency of both the FLCs and the factors is exceptional. The FLCs are most efficient for the longer maturities, and several of the efficiency factors for the $\beta_{k,t}^{(c)}$ exceed one. The slopes $\gamma_k^{(c)}$ are less efficient, but still at least 11\% for all $k,c$.

\begin{table}[ht]
\centering
\begin{tabular}{rrrrr}
\hline
 & $\tau = 8$ & $\tau =90$ & $\tau =180$ & $\tau =270$ \\ 
  \hline
 $f_1(\tau)$ & 0.52 & 0.72 & 0.72 & 0.71 \\ 
$f_2(\tau)$ & 0.48 & 0.72 & 0.73 & 0.71 \\ 
$f_3(\tau)$ & 0.66 & 0.96 & 0.89 & 0.92 \\ 
$f_4(\tau)$ & 0.54 & 0.77 & 0.77 & 0.91 \\ 
$f_5(\tau)$ & 0.61 & 0.72 & 0.84 & 0.85 \\ 
$f_6(\tau)$ & 0.58 & 0.94 & 0.89 & 0.85 \\
   \hline
\end{tabular}
\caption{Efficiency factors for the posterior sampling of  $f_k(\tau), k=1,\ldots,6$, for maturities $\tau \in \{8, 90, 180, 270\}$ months, which are the 2nd, 25th, 50th, and 75th quantiles of the observation points,  using model \eqref{commonHMM} for the yield curve application.}
\end{table}

\begin{table}[ht]
\centering
\begin{tabular}{rrrrrrr}
  \hline
 & 2006-02-10 & 2007-07-06 & 2008-12-05 & 2010-04-30 & 2011-09-23 & 2013-02-22 \\ 
  \hline
$  k = 1,  c = 1 $& 0.83 & 0.91 & 1.00 & 0.83 & 0.91 & 1.00 \\ 
$  k = 2,  c = 1 $& 0.96 & 0.42 & 0.55 & 0.96 & 0.42 & 0.55 \\ 
$  k = 3,  c = 1 $& 0.98 & 0.68 & 1.00 & 0.98 & 0.68 & 1.00 \\ 
$  k = 4,  c = 1 $& 0.91 & 1.00 & 1.01 & 0.91 & 1.00 & 1.01 \\ 
$  k = 5,  c = 1 $& 0.72 & 1.00 & 1.00 & 0.72 & 1.00 & 1.00 \\ 
$  k = 6,  c = 1 $& 0.41 & 0.90 & 1.00 & 0.41 & 0.90 & 1.00 \\ 
$  k = 1,  c = 2 $& 0.95 & 1.00 & 1.00 & 0.95 & 1.00 & 1.00 \\ 
$  k = 2,  c = 2 $& 1.10 & 1.00 & 1.00 & 1.10 & 1.00 & 1.00 \\ 
$  k = 3,  c = 2 $& 0.82 & 1.00 & 1.00 & 0.82 & 1.00 & 1.00 \\ 
$  k = 4,  c = 2 $& 1.04 & 1.02 & 1.00 & 1.04 & 1.02 & 1.00 \\ 
$  k = 5,  c = 2 $& 0.95 & 1.00 & 1.00 & 0.95 & 1.00 & 1.00 \\ 
$  k = 6,  c = 2 $& 1.00 & 1.00 & 1.00 & 1.00 & 1.00 & 1.00 \\ 
$  k = 1,  c = 3 $& 1.00 & 1.00 & 1.00 & 1.00 & 1.00 & 1.00 \\ 
$  k = 2,  c = 3 $& 1.00 & 0.94 & 0.94 & 1.00 & 0.94 & 0.94 \\ 
$  k = 3,  c = 3 $& 1.00 & 1.00 & 1.00 & 1.00 & 1.00 & 1.00 \\ 
$  k = 4,  c = 3 $& 1.00 & 1.00 & 1.00 & 1.00 & 1.00 & 1.00 \\ 
$  k = 5,  c = 3 $& 1.00 & 1.06 & 1.00 & 1.00 & 1.06 & 1.00 \\ 
$  k = 6,  c = 3 $& 1.00 & 1.00 & 0.94 & 1.00 & 1.00 & 0.94 \\ 
$  k = 1,  c = 4 $& 1.00 & 0.95 & 0.96 & 1.00 & 0.95 & 0.96 \\ 
$  k = 2,  c = 4 $& 1.00 & 1.00 & 1.04 & 1.00 & 1.00 & 1.04 \\ 
$  k = 3,  c = 4 $& 1.00 & 1.00 & 1.00 & 1.00 & 1.00 & 1.00 \\ 
$  k = 4,  c = 4 $& 0.92 & 1.00 & 1.00 & 0.92 & 1.00 & 1.00 \\ 
$  k = 5,  c = 4 $& 1.00 & 0.93 & 1.00 & 1.00 & 0.93 & 1.00 \\ 
$  k = 6,  c = 4 $& 1.00 & 1.00 & 1.00 & 1.00 & 1.00 & 1.00 \\ 
   \hline
\end{tabular}
\caption{Efficiency factors for the posterior sampling of  $\beta_{k,t}^{(c)}$ for various times $t$, using model \eqref{commonHMM} for the yield curve application.}
\end{table}

\begin{table}[ht]
\centering
\begin{tabular}{rrrr}
& \multicolumn{3}{c}{Economy} \\
\hline
  & BOE & ECB  & BOC \\ 
  \hline
$k=1$ & 0.44 & 0.39 & 0.15 \\ 
$k=  2$ & 0.12 & 0.11 & 0.12 \\ 
$k=  3$ & 0.40 & 0.38 & 0.19 \\ 
$k=  4$ & 0.42 & 0.26 & 0.19 \\ 
    \hline
\end{tabular}
\caption{Efficiency factors for the posterior sampling of  $\gamma_k^{(c)}$, using model \eqref{commonHMM} for the yield curve application.}
\end{table}

In Figures A.3.1.1, A.3.1.2, and A.3.1.3, we present the trace plots for the FLCs, the factors,  and the slopes, respectively. The vertical gray bars indicate the selected burn-in of 2,000 iterations. Again, the FLCs and the factors demonstrate exceptional MCMC performance. Interestingly, the initializations of the FLCs appear to be farthest from the posterior modes for shorter maturities.  The slopes $\gamma_k^{(c)}$ were initialized at zero, yet congregated around the posterior modes rapidly.

 \begin{figure}[h]
  \centering
\includegraphics[scale= .3]{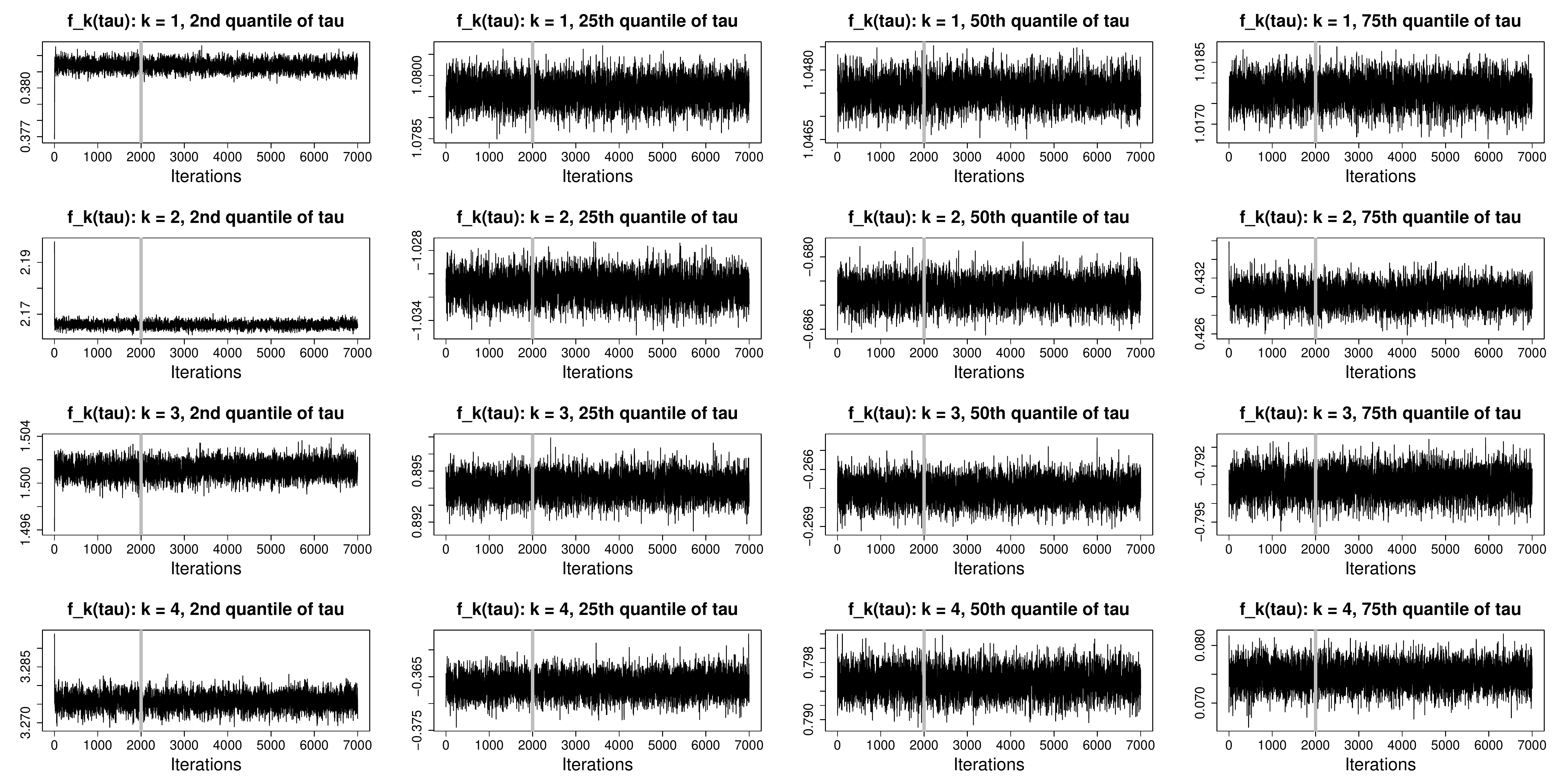} 
  \caption{Trace plots of the posterior samples of $f_k(\tau)$, $k=1,2,3,4$, for the 2nd, 25th, 50th, and 75th quantiles of the observation points, using model \eqref{commonHMM} for the yield curve application. }
\end{figure}

 \begin{figure}[h]
  \centering
\includegraphics[scale= .3]{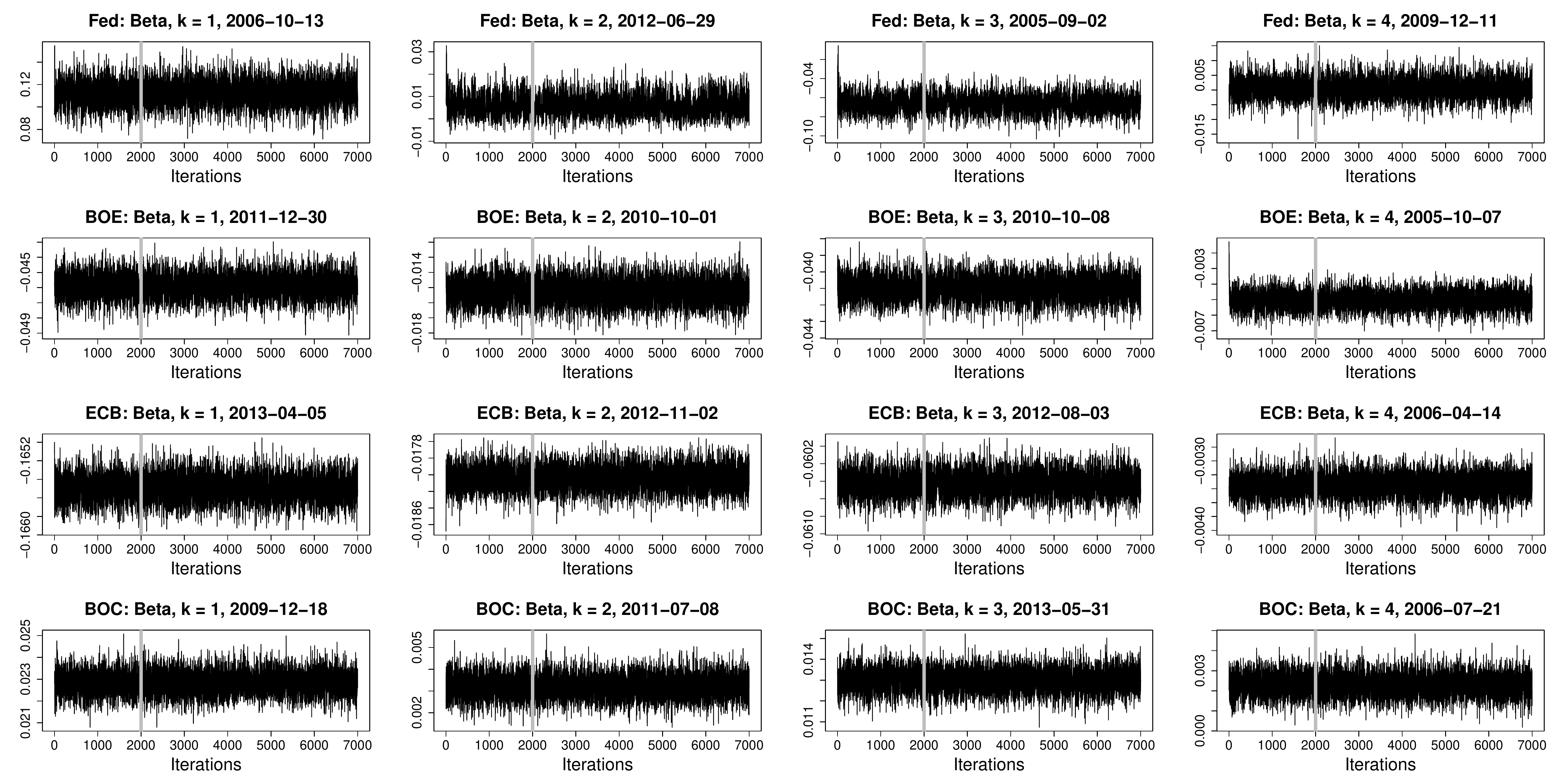} 
  \caption{Trace plots of the posterior samples of $\beta_{k,t}^{(c)}$, $k=1,2,3,4$,  for various times $t$, using model \eqref{commonHMM} for the yield curve application. The vertical gray bar indicates the selected burn-in of $2,000$ iterations.}
\end{figure}

 \begin{figure}[h]
  \centering
\includegraphics[scale= .3]{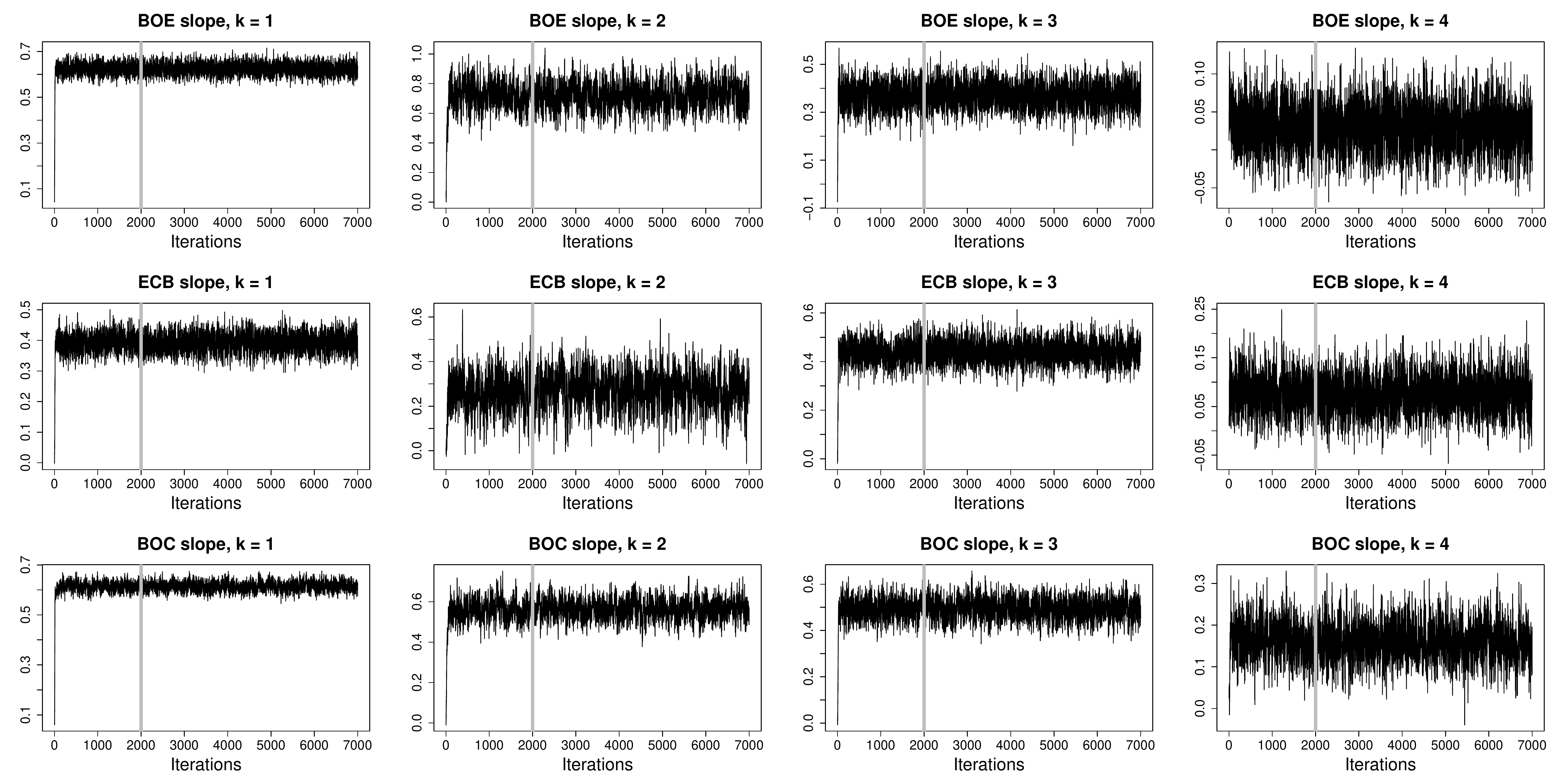} 
  \caption{Trace plots of the posterior samples of $\gamma_k^{(c)}$, $k=1,2,3,4$, using model \eqref{commonHMM} for the yield curve application.}
\end{figure}

\subsubsection{Multivariate Time-Frequency Analysis for Local Field Potential}
We ran the MCMC sampler for $7,000$ iterations and discarded the first $2,000$ iterations as a burn-in. Longer chains and dispersed starting values did not produce noticeably different results.  The sampler was run in \texttt{R}, and took $367$ minutes on a laptop with a 2.40 GHz Intel i7-4700MQ CPU using one core.   

Tables A.3.2.1, and A.3.2.2 contain the efficiency factors for the sample means $\bar{\mu}_t^{(c)}(\tau)$ and the factors $\beta_{k,i,s,t}^{(c)}$ for  various rats $i$, trials $s$, and time bins $t$, respectively. For $\bar{\mu}_t^{(c)}(\tau)$, we compute quantiles of the efficiency factors across all $c,t,\tau$: the minimum efficiency factor is 78\%, while the overwhelming majority of the efficiency factors are at least one. Since we compute pointwise HPD credible intervals for $\bar{\mu}_t^{(c)}(\tau)$ for all $c,t,\tau$, it is encouraging that the MCMC sampler is extremely efficient for these parameters. As in the previous application, the MCMC efficiency of the factors is exceptional. In Figures A.3.2.1 and A.3.2.2, we present the trace plots for   $\bar{\mu}_t^{(c)}(\tau)$  and $\beta_{k,i,s,t}^{(c)}$. The MCMC performance for both sets of parameters appears to be very good.

\begin{table}[ht]
\centering
\begin{tabular}{cccccc}
Min. & 25th Quantile & Median & Mean & 75th Quantile  & Max. \\
\hline
 0.7781 &  1.0000  &1.0000 &  1.0060 &  1.0000  &1.8270\\
  \hline
\end{tabular}
\caption{Summary statistics of the efficiency factors for the posterior sampling of $\bar{\mu}_t^{(c)}(\tau)$ across all $c, t, \tau$,  using model \eqref{ratmod} for the LFP application.}
\end{table}
\begin{table}[ht]
\centering
\begin{tabular}{rrrrrrr}
  \hline
 & $720$ & $1440$ & $2160$ & $2880$ & $3600$ & $4320$ \\ 
  \hline
$k = 1,  c = 1$ & 1.00 & 1.00 & 1.00 & 1.00 & 1.00 & 1.00 \\ 
$  k = 2,  c = 1$ & 1.00 & 0.90 & 1.06 & 1.03 & 1.00 & 1.09 \\ 
$  k = 3,  c = 1$ & 1.00 & 1.22 & 1.00 & 1.00 & 0.99 & 1.07 \\ 
$  k = 4,  c = 1$ & 1.00 & 1.00 & 1.00 & 1.00 & 1.00 & 1.08 \\ 
$  k = 5,  c = 1$ & 1.00 & 1.00 & 1.00 & 1.05 & 1.00 & 1.00 \\ 
$  k = 6,  c = 1$ & 1.00 & 0.90 & 1.00 & 1.11 & 0.98 & 1.00 \\ 
$  k = 7,  c = 1$ & 0.93 & 1.00 & 1.00 & 1.00 & 1.00 & 1.00 \\ 
$  k = 8,  c = 1$ & 1.00 & 1.10 & 1.00 & 0.94 & 1.00 & 1.00 \\ 
$  k = 9,  c = 1$ & 1.00 & 1.00 & 1.00 & 1.00 & 1.00 & 1.00 \\ 
$  k = 10,  c = 1$ & 1.00 & 1.00 & 1.00 & 1.00 & 0.96 & 1.00 \\ 
$  k = 1,  c = 2$ & 1.00 & 1.00 & 1.00 & 1.00 & 1.10 & 1.00 \\ 
$  k = 2,  c = 2$ & 0.94 & 1.05 & 1.00 & 1.00 & 1.00 & 1.00 \\ 
$  k = 3,  c = 2$ & 1.00 & 1.07 & 1.00 & 1.00 & 0.87 & 0.94 \\ 
$  k = 4,  c = 2$ & 1.13 & 1.00 & 1.00 & 1.01 & 0.95 & 0.89 \\ 
$  k = 5,  c = 2$ & 1.00 & 1.00 & 1.00 & 1.00 & 0.95 & 1.00 \\ 
$  k = 6,  c = 2$ & 1.00 & 1.12 & 1.00 & 1.05 & 1.01 & 1.00 \\ 
$  k = 7,  c = 2$ & 1.00 & 1.06 & 1.00 & 1.00 & 1.00 & 1.00 \\ 
$  k = 8,  c = 2$ & 1.00 & 1.14 & 1.05 & 1.00 & 1.07 & 1.00 \\ 
$  k = 9,  c = 2$ & 0.88 & 1.00 & 0.95 & 1.00 & 1.00 & 1.00 \\ 
$  k = 10,  c = 2$ & 1.00 & 1.03 & 1.00 & 1.00 & 1.00 & 1.00 \\ 
$  k = 1,  c = 3$ & 1.00 & 1.07 & 1.00 & 1.00 & 1.15 & 0.95 \\ 
$  k = 2,  c = 3$ & 1.00 & 1.07 & 1.00 & 0.95 & 1.00 & 0.90 \\ 
$  k = 3,  c = 3$ & 1.00 & 1.00 & 1.00 & 0.95 & 1.00 & 1.00 \\ 
$  k = 4,  c = 3$ & 1.00 & 1.00 & 0.93 & 1.06 & 1.00 & 1.00 \\ 
$  k = 5,  c = 3$ & 1.00 & 1.00 & 1.00 & 1.00 & 1.00 & 1.00 \\ 
$  k = 6,  c = 3$ & 1.00 & 1.00 & 0.94 & 0.97 & 1.00 & 1.00 \\ 
$  k = 7,  c = 3$ & 1.00 & 1.00 & 0.95 & 1.00 & 1.00 & 1.00 \\ 
$  k = 8,  c = 3$ & 1.00 & 1.00 & 1.00 & 0.86 & 0.95 & 1.00 \\ 
$  k = 9,  c = 3$ & 1.15 & 1.00 & 1.00 & 1.00 & 1.00 & 1.00 \\ 
$  k = 10,  c = 3$ & 1.04 & 1.00 & 1.00 & 1.00 & 1.00 & 1.00 \\ 
   \hline
\end{tabular}
\caption{Efficiency factors for the posterior sampling of  $\beta_{k,i,s,t}^{(c)}$, using model \eqref{ratmod} for the LFP application. The column indexes are the 15th, 30th, 45th, 60th, 75th, and 90th quantiles of 1:4800, which is the concatenated time index across rats $i=1,\ldots,8$, trials $s=1,\ldots,40$, and time bins $t=1,\ldots, 15$. }
\end{table}

  \begin{figure}[h]
  \centering
\includegraphics[scale= .28]{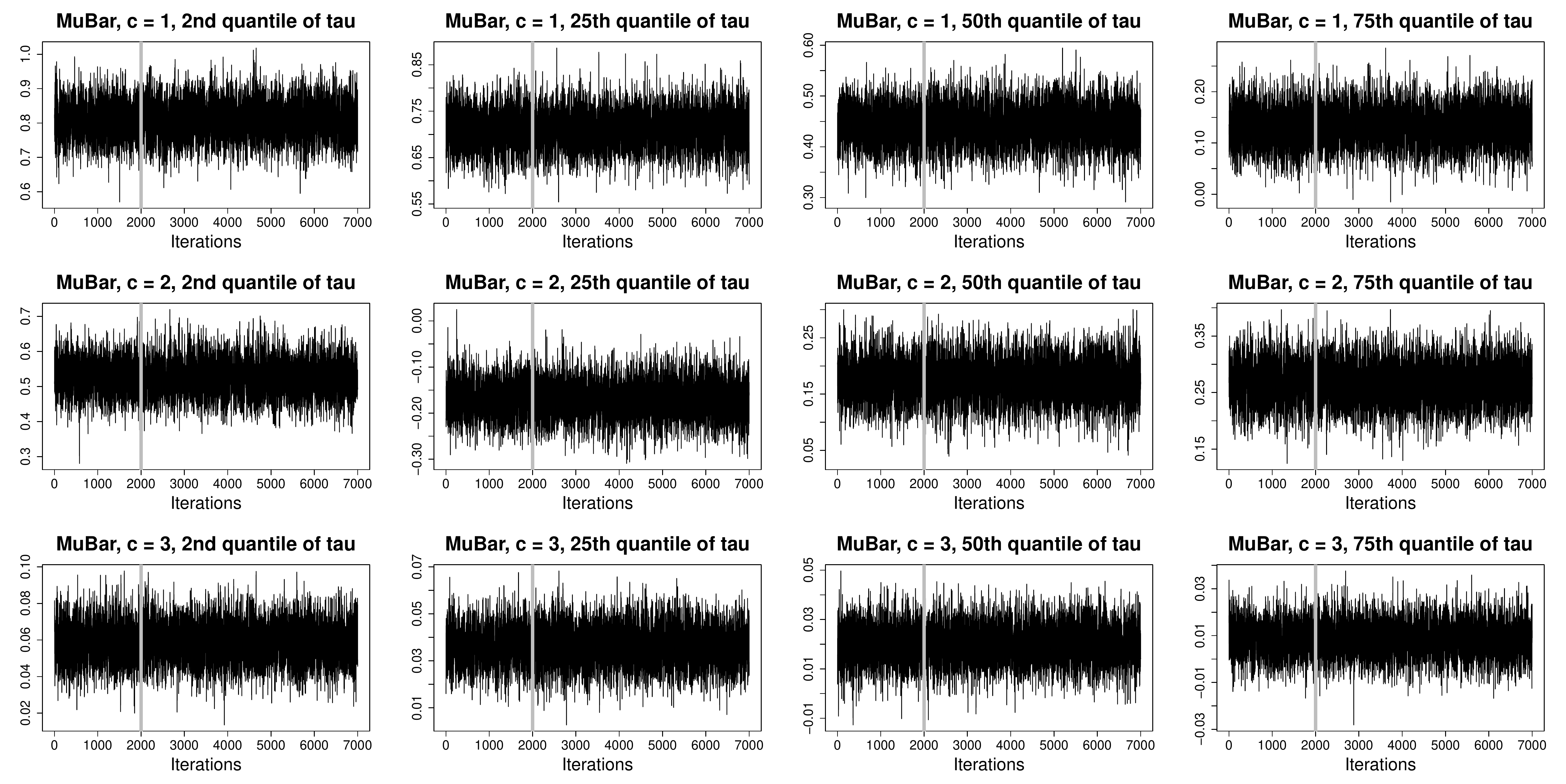} 
  \caption{Trace plots of the posterior samples of $\bar{\mu}_t^{(c)}(\tau)$, for the 2nd, 25th, 50th, and 75th quantiles of the observation points, $c=1,\ldots,C$, and selected time bins, using model \eqref{ratmod} for the LFP application. The vertical gray bar indicates the selected burn-in of $2,000$ iterations.}
\end{figure}
 \begin{figure}[h]
  \centering
\includegraphics[scale= .32]{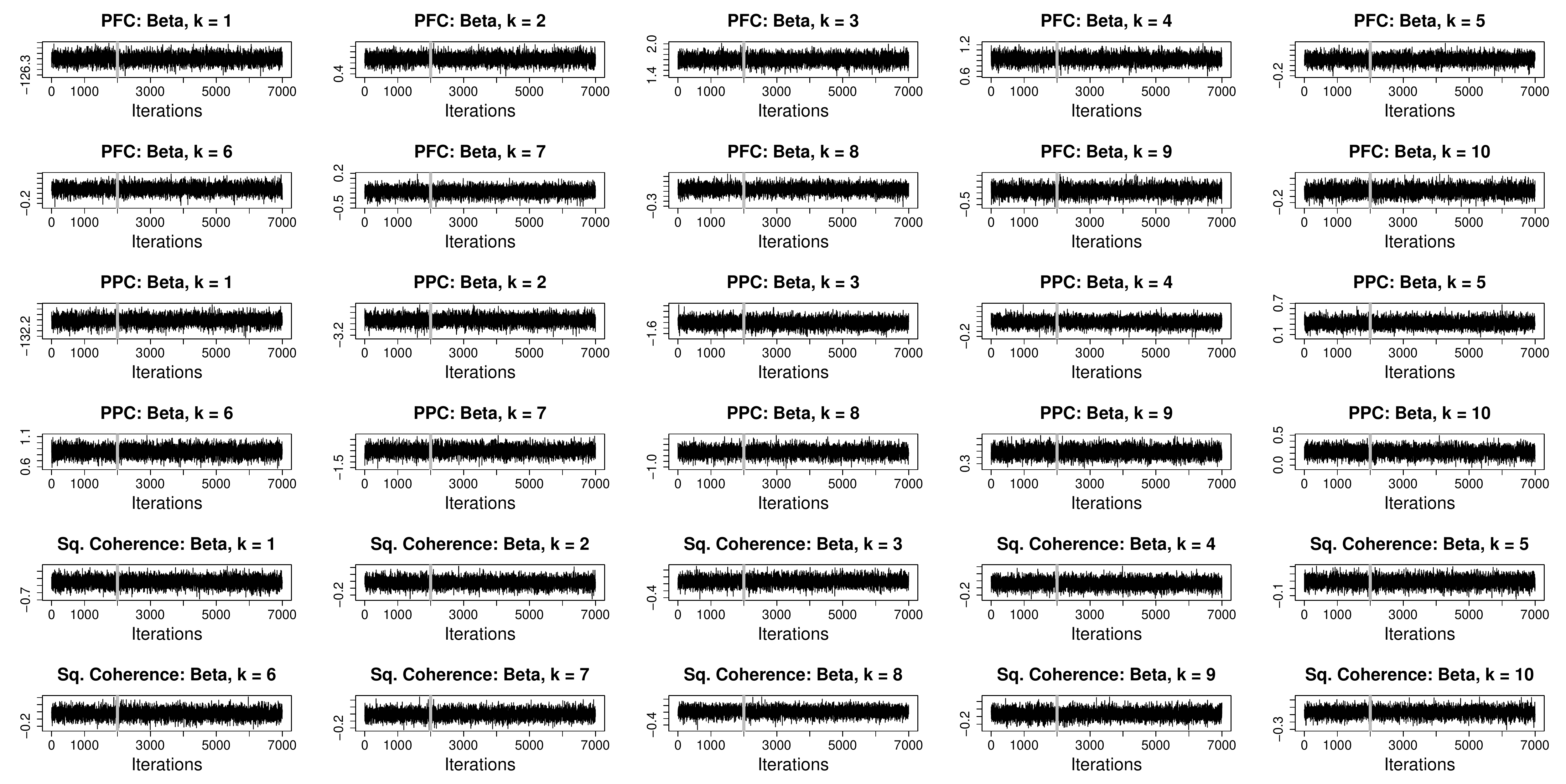} 
  \caption{Trace plots of the posterior samples of $\beta_{k,i,s,t}^{(c)}$  for various $(i,s,t)$, using model \eqref{ratmod} for the LFP application.}
\end{figure}

\subsection{The Common Trend Hidden Markov Model}\label{commontrend}
Consider the following extension of the common trend model \eqref{commonHMM} in the main paper:
\begin{equation}\label{commonHMM2}
\begin{cases}
\beta_{k,t}^{(1)} = \omega_{k,t}^{(1)}\\
\beta_{k,t}^{(c)} =  s_{k,t}^{(c)} (\gamma_k^{(c)}\beta_{k,t}^{(1)}) + \omega_{k,t}^{(c)} & c=2,\ldots,C
\end{cases}
\end{equation}
where $\left\{s_{k,t}^{(c)}: t=1,\ldots,T\right\}$ is a discrete Markov chain with states $\{0,1\}$.  Model \eqref{commonHMM2} reduces to model \eqref{commonHMM} in the main paper when  $s_{k,t}^{(c)} = 1$ for all $k,c,t$. As with the common trend model, we can use \eqref{commonHMM2} to  investigate how the   factors $ \beta_{k,t}^{(c)}$ for each economy $c>1$ are {\it directly} related to those of the Fed, $ \beta_{k,t}^{(1)}$.  Model \eqref{commonHMM2} relates each economy $c>1$ to the Fed using a  regression framework, in which we regress $\beta_{k,t}^{(c)}$ on $\beta_{k,t}^{(1)}$  with AR($r$) errors, where the (Fed) predictor $ \beta_{k,t}^{(1)}$ is present at time $t$ only if $s_{k,t}^{(c)}=1$. Therefore, the role of the  states $s_{k,t}^{(c)}$ is to identify times $t$ for which $ \beta_{k,t}^{(c)}$ is strongly correlated with $ \beta_{k,t}^{(1)}$; i.e., the periods for which the week-to-week changes in the features of the yield curves described by $f_k$ are similar for economy $c$ and the Fed. When $s_{k,t}^{(c)} = s_{k,t}^{(c')} = 1$ for $c\ne c'$, we also have dependence between $ \beta_{k,t}^{(c)}$ and $ \beta_{k,t}^{(c')}$; therefore, in \eqref{commonHMM2}, the Fed acts as a conduit for {\it all} contemporaneous dependence between economies.

It is natural for the values of the states $s_{k,t}^{(c)}$ to depend on past values of the states: if $ \beta_{k,t}^{(c)}$ is correlated with $ \beta_{k,t}^{(1)}$ at time $t$, then we may perhaps infer something about their relative behavior at time $t+1$. Following the construction of \cite{albert1993bayes}, the  distribution  of $\left\{ s_{k,t}^{(c)}: t=1,\ldots,T\right\}$, unconditional on the factors $\beta_{k,t}^{(c)}$, is determined by $ P(s_{k,t}^{(c)} = 1 | s_{k,t-1}^{(c)}=0)  = q_{01,k}^{(c)}$ and $P(s_{k,t}^{(c)} = 0 | s_{k,t-1}^{(c)}=1) = q_{10,k}^{(c)}$
with the accompanying Markov property $\left[s_{k,t}^{(c)} \big| s_{k,t-1}^{(c)}, s_{k,t-2}^{(c)}, \ldots\right] = \left[s_{k,t}^{(c)} \big| s_{k,t-1}^{(c)}\right]$, where the transition probabilities $q_{01,k}^{(c)}$ and $q_{10,k}^{(c)}$ are unknown. Therefore, \eqref{commonHMM2} contains a {\it hidden Markov model}, where the hidden states $s_{k,t}^{(c)}$ determine whether or not the  factors $ \beta_{k,t}^{(c)}$ are related to those of the Fed, $ \beta_{k,t}^{(1)}$, at time $t$. As in \cite{albert1993bayes}, we use conjugate Beta priors for the transition probabilities, and select the hyperparameters so that the bulk of the mass of the prior distribution is on $(0, 0.5)$, which reflects the belief that transitions should occur infrequently.  Sampling from the posterior distribution of $\left\{ s_{k,t}^{(c)}: t=1,\ldots,T\right\}$ (i.e., conditional on the factors $\beta_{k,t}^{(c)}$) is a straightforward application of  \cite{albert1993bayes}.



\subsubsection{Sampling The Common Trend Hidden Markov Model}\label{samplecommontrend}
 While model \eqref{commonHMM2}  is a useful example of the flexibility of the MFDLM, it is not supported by DIC: the DIC for model \eqref{commonHMM} is $-2,393,266$, while the DIC for model \eqref{commonHMM2} is $-2,393,200$.   However, since we can obtain the preferred model \eqref{commonHMM} from the main paper by setting $s_{k,t}^{(c)} = 1$, we describe the DLM construction for the more general model \eqref{commonHMM2}. Expressing \eqref{commonHMM2} as a DLM allows us to use efficient state space samplers for the factors $\utwi{\beta}_t$, as in the algorithm described in Section \ref{algorithm}.

We can express \eqref{commonHMM2} as the $\utwi{\beta}_t = \utwi{\theta}_t$-level in \eqref{fhdlm} with $\mathbf{X}_t = \mathbf{I}_{CK\times CK}$ and  $\mathbf{V}_t = \mathbf{0}_{CK\times CK}$.  Let $ \utwi{L_{\beta}}_t = \mathbf{I}_{CK\times CK} - \mathbf{Q}_{t}$,  
$$
\mathbf{Q}_{t} =  \begin{pmatrix} \mathbf{0}_{K\times K} &  \mathbf{0}_{K\times K} & \cdots & \mathbf{0}_{K\times K} \\  \mathbf{S}_{t}^{(2)} \utwi{\gamma}^{(2)} &  \mathbf{0}_{K\times K} & \cdots & \mathbf{0}_{K\times K} \\ \vdots & \vdots & \ddots & \vdots \\ \mathbf{S}_{t}^{(C)}  \utwi{\gamma}^{(C)}&  \mathbf{0}_{K\times K} & \cdots & \mathbf{0}_{K\times K}
 \end{pmatrix},
 $$
 where $\mathbf{S}_{t}^{(c)} = \mbox{diag}(\{s_{k,t}^{(c)}\}_{k=1}^K)$ and $ \utwi{\gamma}^{(c)}= \mbox{diag}(\{\gamma_{k}^{(c)}\}_{k=1}^K)$. Note that $\utwi{L_{\beta}}_t^{-1} = \mathbf{I}_{CK\times CK} + \mathbf{Q}_{t}$. In vector notation, \eqref{commonHMM2} can be written
 \begin{equation}
\utwi{L_{\beta}}_t \utwi{\beta}_t = \utwi{\Psi} \utwi{L_{\beta}}_{t-1} \utwi{\beta}_{t-1} + \utwi{\tilde \omega}_t
\end{equation}
where $\utwi{\Psi} = \mbox{diag}(\{\psi_{k,1}^{(c)}\}_{k,c})$   and $\utwi{\tilde \omega}_t$ has elements $\tilde{\omega}_{k,t}^{(c)} = \sigma_{k,(c),t} z_{k,t}^{(c)}$ with $\utwi{\tilde \omega}_t \sim N(\utwi{0}, \mathbf{\tilde W}_t)$ and $\mathbf{\tilde W}_t = \mbox{diag}(\{\sigma_{k, (c),t}^2\}_{k,c})$ 
.  Inverting $\utwi{L_{\beta}}_t$, the DLM evolution equation is therefore  
 \begin{equation}
\utwi{\beta}_t  = \mathbf{G}_t \utwi{\beta}_{t-1} + \utwi{ \omega}_t \\
\end{equation}
where $\mathbf{G}_t =  (\mathbf{I}_{CK\times CK} + \mathbf{Q}_{t}) \utwi{\Psi}  (\mathbf{I}_{CK\times CK} - \mathbf{Q}_{t-1})$ and $\utwi{\omega}_t =  (\mathbf{I}_{CK\times CK} + \mathbf{Q}_{t}) \utwi{\tilde \omega}_t \sim N(\utwi{0}, \mathbf{W}_t)$, with $\mathbf{W}_t = \utwi{L_{\beta}}_t^{-1}\mathbf{\tilde W}_t (\utwi{L_{\beta}}_t^{-1})'.$
Since $\mathbf{Q}_t \utwi{\Psi} \mathbf{Q}_{t-1} = \utwi{0}_{CK \times CK}$, we have
$$
\mathbf{G}_t = \begin{pmatrix} \utwi{\Psi}^{(1)} &  \mathbf{0}_{K\times K} & \cdots & \mathbf{0}_{K\times K} \\  \utwi{\gamma}^{(2)}\left(\mathbf{S}_{t}^{(2)} \utwi{\Psi}^{(1)} -\mathbf{S}_{t-1}^{(2)} \utwi{\Psi}^{(2)} \right)&  \utwi{\Psi}^{(2)} & \cdots & \mathbf{0}_{K\times K} \\ \vdots & \vdots & \ddots & \vdots \\ \utwi{\gamma}^{(C)}\left(\mathbf{S}_{t}^{(C)} \utwi{\Psi}^{(1)} -\mathbf{S}_{t-1}^{(C)} \utwi{\Psi}^{(C)} \right)&  \mathbf{0}_{K\times K} & \cdots &\utwi{\Psi}^{(C)}
 \end{pmatrix},
 $$
where $\utwi{\Psi}^{(c)} = \mbox{diag}(\{\psi_{k,1}^{(c)}\}_k)$. Similarly, we may compute $\mathbf{W}_t =   (\mathbf{I}_{CK\times CK} + \mathbf{Q}_{t})\mathbf{\tilde W}_t (\mathbf{I}_{CK\times CK} + \mathbf{Q}_{t}') = \mathbf{\tilde W}_t + \mathbf{Q}_t\mathbf{\tilde W}_t + (\mathbf{Q}_t\mathbf{\tilde W}_t)' + \mathbf{Q}_t\mathbf{\tilde W}_t\mathbf{Q}_t'$. Letting $\utwi{\sigma}_{(c),t}^2  = \mbox{diag}(\{\sigma_{k,(c),t}^2\}_{k=1}^K)$  so that $\mathbf{\tilde W}_t = \mbox{bdiag}(\utwi{\sigma}_{(1),t}^2, \ldots, \utwi{\sigma}_{(C),t}^2)$, we may compute the relevant terms explicitly:
$$
\mathbf{Q}_t \mathbf{\tilde W}_t = \begin{pmatrix} \mathbf{0}_{K\times K} & \mathbf{0}_{K\times K} & \cdots & \mathbf{0}_{K\times K} \\ \mathbf{S}_t^{(2)} \utwi{\gamma}^{(2)} \utwi{\sigma}_{(1),t}^2 & \mathbf{0}_{K\times K} & \cdots & \mathbf{0}_{K\times K}\\
\vdots & \vdots & \ddots & \vdots \\
\mathbf{S}_t^{(C)} \utwi{\gamma}^{(C)} \utwi{\sigma}_{(1),t}^2 & \mathbf{0}_{K\times K} & \cdots & \mathbf{0}_{K\times K}
\end{pmatrix} $$
and 
$$
\mathbf{Q}_t \mathbf{\tilde W}_t\mathbf{Q}_t'  = \begin{pmatrix}   \mathbf{0}_{K\times K} & \mathbf{0}_{K\times K} & \cdots & \mathbf{0}_{K\times K} \\ 
\mathbf{0}_{K\times K} &\mathbf{S}_t^{(2)} \utwi{\gamma}^{(2)} \utwi{\sigma}_{(1),t}^2 \mathbf{S}_t^{(2)} \utwi{\gamma}^{(2)}& \cdots  &\mathbf{S}_t^{(2)} \utwi{\gamma}^{(2)} \utwi{\sigma}_{(1),t}^2 \mathbf{S}_t^{(C)} \utwi{\gamma}^{(C)}\\
\vdots & \vdots & \ddots & \vdots \\
\mathbf{0}_{K\times K} &\mathbf{S}_t^{(C)} \utwi{\gamma}^{(C)} \utwi{\sigma}_{(1),t}^2 \mathbf{S}_t^{(2)} \utwi{\gamma}^{(2)}  & \cdots & \mathbf{S}_t^{(C)} \utwi{\gamma}^{(C)} \utwi{\sigma}_{(1),t}^2 \mathbf{S}_t^{(C)} \utwi{\gamma}^{(C)}
\end{pmatrix} 
$$
where again, the component terms are all diagonal, and therefore can be reordered for convenience. Combining  terms and simplifying, the  error variance matrix is 
$$
\mathbf{W}_t = \begin{pmatrix}    \utwi{\sigma}_{(1),t}^2  & \mathbf{S}_t^{(2)} \utwi{\gamma}^{(2)} \utwi{\sigma}_{(1),t}^2 & \cdots & \mathbf{S}_t^{(C)} \utwi{\gamma}^{(C)} \utwi{\sigma}_{(1),t}^2 \\ 
 \mathbf{S}_t^{(2)} \utwi{\gamma}^{(2)} \utwi{\sigma}_{(1),t}^2 &  \utwi{\sigma}_{(2),t}^2 +  \mathbf{S}_t^{(2)} (\utwi{\gamma}^{(2)})^2 \utwi{\sigma}_{(1),t}^2 & \cdots  &\mathbf{S}_t^{(2)} \mathbf{S}_t^{(C)} \utwi{\gamma}^{(2)}\utwi{\gamma}^{(C)} \utwi{\sigma}_{(1),t}^2 \\
\vdots & \vdots & \ddots & \vdots \\
 \mathbf{S}_t^{(C)} \utwi{\gamma}^{(C)} \utwi{\sigma}_{(1),t}^2&\mathbf{S}_t^{(2)} \mathbf{S}_t^{(C)} \utwi{\gamma}^{(2)}\utwi{\gamma}^{(C)} \utwi{\sigma}_{(1),t}^2  & \cdots &\utwi{\sigma}_{(C),t}^2 +  \mathbf{S}_t^{(C)} (\utwi{\gamma}^{(C)})^2 \utwi{\sigma}_{(1),t}^2 
\end{pmatrix}.
$$
When $s_{k,t}^{(c)} = 1, c>1$ the slope parameter  $\gamma_k^{(c)}$ may increase or decrease the error variance of the residuals $\tilde \omega_{k,t}^{(c)}$ at time $t$, and determines the contemporaneous covariance between $\tilde \omega_{k,t}^{(c)}$ and $\tilde \omega_{k,t}^{(1)}$. Similarly, when $s_{k,t}^{(c)} = s_{k,t}^{(c')} = 1$, the product   $\gamma_k^{(c)}\gamma_k^{(c')}\sigma_{k,(1),t}^2$  determines the contemporaneous covariance between $\tilde\omega_{k,t}^{(c)}$ and $\tilde\omega_{k,t}^{(c')}$ at time $t$. 


\numberwithin{figure}{subsection}

\subsection{Additional Figures}
\begin{sidewaysfigure}[h]
  \centering
\includegraphics[scale= .45]{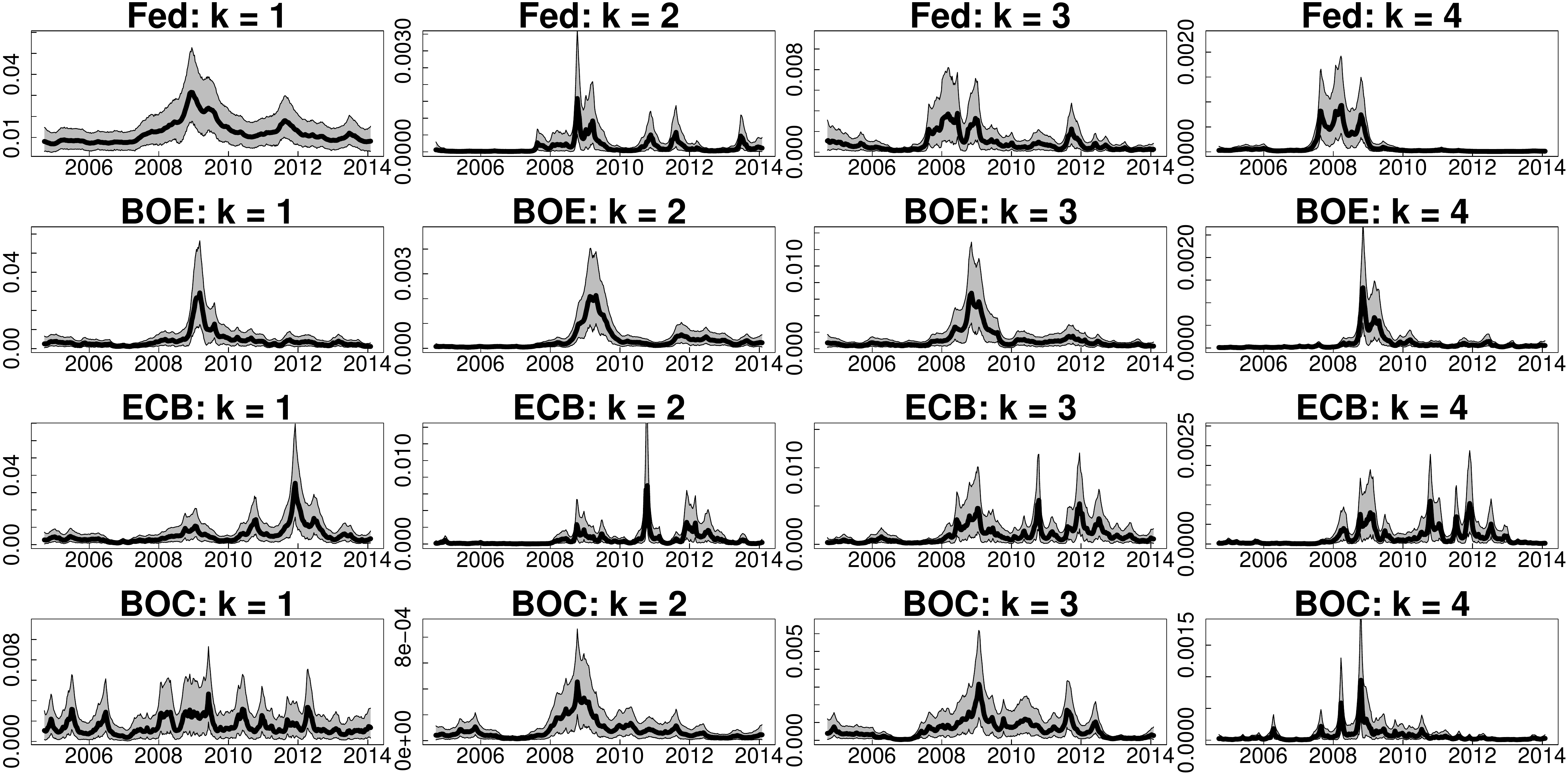} 
  \caption{Posterior means (black line) and 95\% HPD intervals (gray shading) of the volatilities $\sigma_{k,(c),t}^2$ from model \eqref{commonHMM} in the main paper.}
\end{sidewaysfigure}

\begin{figure}[h]
  \centering
\includegraphics[scale= .24]{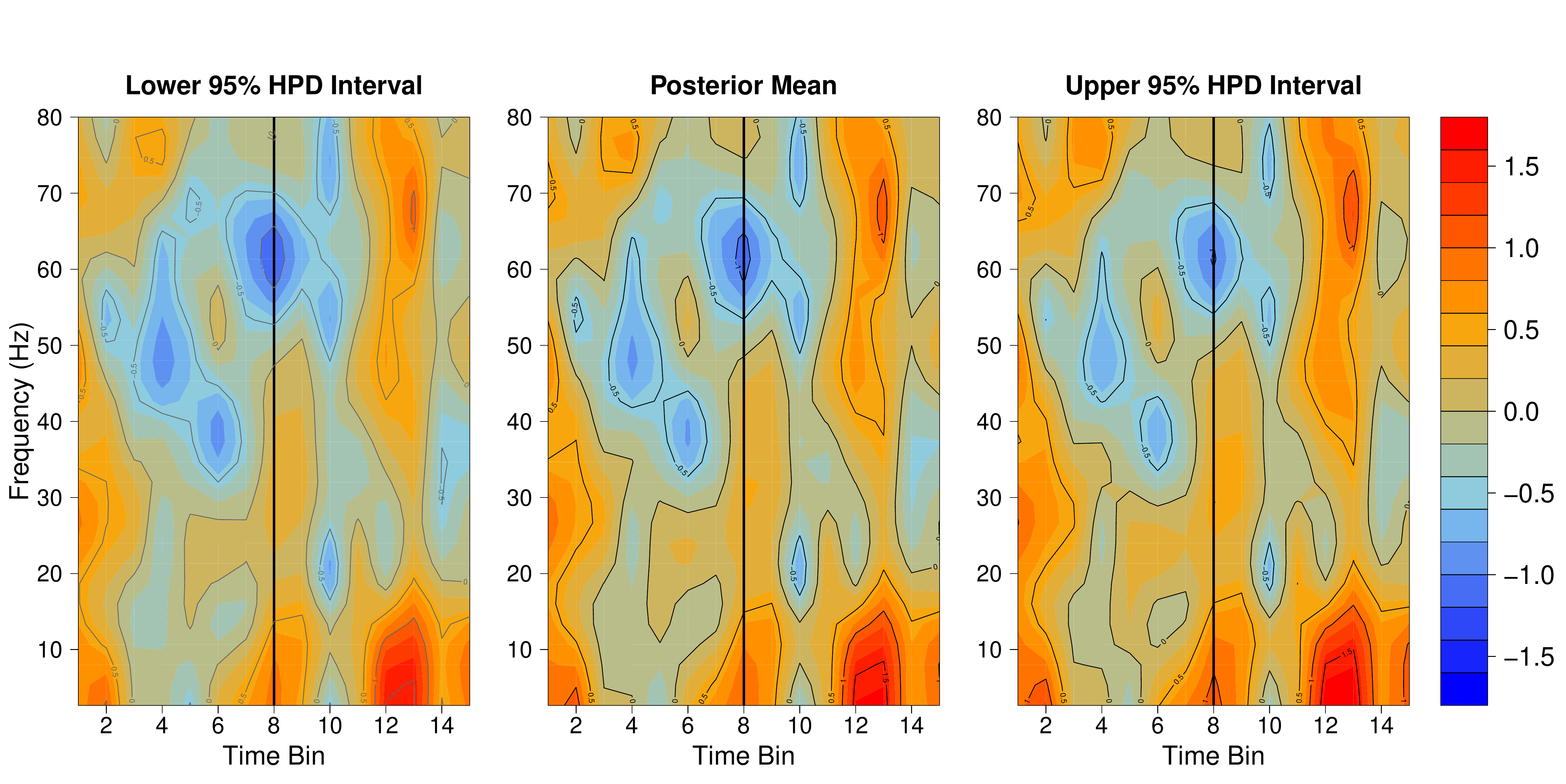} 
  \caption{Pointwise 95\% HPD intervals and the posterior mean for $\bar{\mu}_t^{(1)}$, which is the average difference in the PFC log-spectra between the FC and FS trials. The black vertical lines indicate  the event time $t^*$.}
\includegraphics[scale= .24]{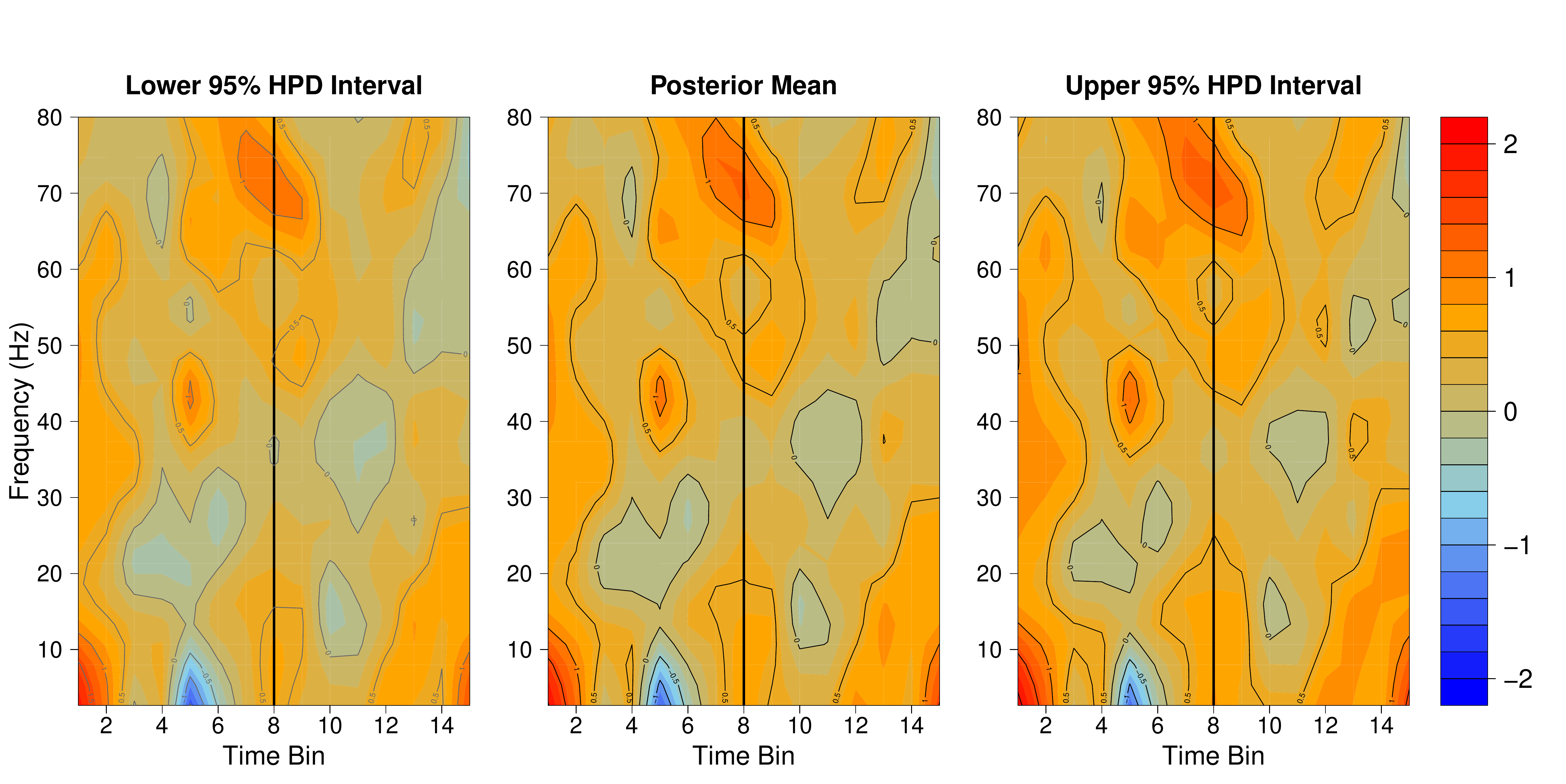} 
  \caption{Pointwise 95\% HPD intervals and the posterior mean for $\bar{\mu}_t^{(2)}$, which is the average difference in the PFC log-spectra between the FC and FS trials. The black vertical lines indicate  the event time $t^*$.}
      \includegraphics[scale= .24]{coh.pdf} 
  \caption{Pointwise 95\% HPD intervals and the posterior mean for $\bar{\mu}_t^{(3)}$, which is the average difference in squared coherence between the FC and FS trials. The black vertical lines indicate the event time  $t^*$.}
\end{figure}

\end{document}